\documentclass[twocolumn,aps,floatfix,showpacs,prd,longbibliography]{revtex4-2}
\usepackage{graphics, epsfig ,color} % figures en eps
\usepackage{amsmath,amssymb} % environnement American Mathematical Society
\usepackage{amsfonts,subfigure}
\usepackage{ulem}
\usepackage{enumerate}
\usepackage[dvipsnames]{xcolor}

\usepackage{xcolor, soul}
\sethlcolor{yellow}

\graphicspath{{figures/}}

\begin{document}
\date{\today}
\title{Optimally frequency synchronized networks of non-identical Kuramoto oscillators }
\author{L. Gil }
\affiliation{Universit\'e C\^ote d'Azur, Institut de Physique de Nice (INPHYNI), France}

\begin{abstract}
Based on a local greedy numerical algorithm, we compute the topology of weighted, directed, and of unlimited extension networks of non-identical Kuramoto oscillators which simultaneously satisfy 2 criteria:  i) global frequency synchronization  and ii) minimum total mass of the connection weights. This problem has been the subject of many previous interesting studies, but from our best knowledge, this is the first time that no a priori constraint is imposed, either on the form or on the dynamics of the connections. The results are surprising: the optimal networks turn out to be strongly symmetric, very economical, to display a strong rich club structure, and next to the already reported strong correlation between natural frequencies and the weight of incoming connections, we also observe a correlation, even more marked, between these same natural frequencies and the weight of outgoing connections. The latter result is at odds with theoretical predictions.
\end{abstract}
 \pacs{89.75.-k Complex systems - 89.75.Fb Structures and organization of complex systems - 05.45.Xt Synchronization; coupled oscillators -  05.65.+b Self-organized systems}
\maketitle

\section{Introduction}
Numerous complex systems in biology, society, economics and technology involve many interacting units organized as adaptative networks in order to collectively process information, electricity, matter or energy. Mycelial fungi and acellular slime molds,  plant circadian oscillation, blood vessel growth, genetic and neural networks, internet, road and rail networks, mail and collaborators, power grids, opinion and disease spreading constitute a non-exhaustive list of examples that have been reported on  by an abundant literature \cite{AdaptativeNetworksReviews1,AdaptativeNetworksReviews2,AdaptativeNetworksReviews3}. 

Besides their exceptional ubiquity, adaptative networks display also fascinating features: although based on simple local rules, they can self-organize robustly toward phase transitions \cite{Bornholdt2000,Bornholdt2003,Garlaschelli2007} and highly non-trivial complex topologies involving modular structure and cluster synchronization \cite{Zanette2006,Gutierrez2011}. As evidence of the adaptation of these networks, strong correlations between frequency and incoming connection weight distributions are reported \cite{Brede2008a,Brede2008b,Papadopoulos2017}. The resulting increase in synchronization proves that the adaptative network is able to make the most of the heterogeneity of its components \cite{Nashikawa2016}.

From a theoretical point of view, Kuramoto's model is a well known, omnipresent and useful paradigmatic framework to study synchronization of non-identical oscillators \cite{Strogatz2000,Acebron2005}. In a first step, this model has been extended to heterogeneous but {\it stationary} network topologies to study the influence of scale-free degree distribution \cite{Moreno2004,Lee2005,GomezPRE2007,GomezPRL2007}, small world architecture \cite{Hong2002} or community structure onto the synchronization \cite{Hong2002,Oh2005,Perez2006}. Then in a second step, various adaptive Kuramoto's models have been proposed in order to explore the effects of co-evolution and adaptation. In section \ref{Old Kuramoto} we review these models and notice that they mainly fall into 2 categories: On the one hand, "bio-inspired" ones \cite{Seliger2002,Timms2014,Zhao2007,Assenza2011,AssenzaPRL2011,Zanette2006,Papadopoulos2017,Eom2016,Rentzeperis2020,Avalos2018} for which the network adaptation rules are guessed from biological analogies and fixed in advance. Their purpose is to verify that the rules inspired by biology actually lead to the observed behaviours. On the other hand, the "random rewired" models \cite{Brede2008a, Brede2008b}, which do not presuppose anything on the evolution rules but only retain those random rewiring which do correspond to an increase in the synchronization order parameter. Under fairly strong hypotheses dealing both with the non-directed aspect (ND) of the network, constant total mass of the connection weight (CTM), existence of a pre-set upper limit (PSUL) for the weight of the connections, restriction to only 2 values (2V) of the possible weights of the connections, these models predict the formation of modular structures but only transient, as well as the occurence of steady strong correlation between the natural frequency distribution and those of the incoming connection weight.
 
 In this study, we revisit the conditions of full synchronization of a finite set of non-identical Kuramoto oscillators by releasing as many constraints as possible either on the form or on the dynamics of their connection network. Thus we numerically compute  the topology of non-identical Kuramoto oscillator networks which are both weighted, directed, and of unlimited extension (i.e. the maximal value of the weight is not limited a priori). We only require that the oscillators are completely synchronized and that the economic cost of the network, i.e. the total mass of the connection weights, is locally as low as possible. 
The paper is organized as follows: section \ref{Old Kuramoto} is devoted to a review of the previous adaptative Kuramoto's models, section \ref{Algorithm} deals with the presentation of the numerical algorithm, finally the results are shown in section \ref{Results} and discussed in the last section \ref{Conclusion}.

\section{Previous adaptative Kuramoto's model}
\label{Old Kuramoto}
To describe and classify the various adaptation rules that have been investigated, we begin by introducing the mathematical notation that is common to all of them. The Kuramoto's models deal with a set of $N$ non-identical phase oscillators whose dynamics are ruled by
\begin{equation}
\partial_{t} \theta_{i}=\omega_{i} +{{1}\over{N}} \displaystyle{\sum _{j=1}^{N} W_{ij} sin(\theta_{j}-\theta_{i})}
\label{KuramotoEquation}
\end{equation}
where $\theta_{i}$ is the phase of the ith oscillator, $\omega_{i}$ its natural frequency and $W_{ij}$ stands for the influence of $j$ onto $i$. In the original model the oscillators are all to all, homogeneously and statically connected. But in the case of adaptive networks, the whole game consists in finding an interesting dynamics for the weight of the connections. Appendix \ref{Appendix} provides a brief presentation of the different dynamics that have been previously considered, with an emphasis on the underlying assumptions. To summarise, two main sources of inspiration can be identified. Some adaptative models \cite{Seliger2002,Timms2014,Zhao2007,Assenza2011,AssenzaPRL2011,Avalos2018} are clearly  "bio inspired". The dynamics of the weight is deterministic, described once for all by an ordinary differential equation, mimicking well known biological rules of adaptation like Hebbian, anti-Hebbian, Spike-timing-dependent plasticity (STDP) or homeostasis. 
Models \cite{Zanette2006,Papadopoulos2017,Eom2016,Rentzeperis2020} are also "bio inspired", but the rewiring dynamics is discretized. At each rewiring step, a node $i$ is randomly chosen. Then a link between $i$ and the node $j_{1}$ is substituted with a link between $i$ and $j_{2}$ accordingly to the relative fitness of the links.  The fitness is defined on the basis of expected biological adaptation rules. Finally \cite{Brede2008a, Brede2008b} are of pure "random-wired" type. At each time step, the proposed random rewiring of the network is validated only if it leads to a decrease of the synchronization order parameter.
The "bio inspired"  models compute the consequences of the biological rules onto the network topology and dynamics. Synchronization is not explicitly imposed and it is already an important result to show the existence of a parameter regime where it can be observed. The observation of characteristic properties like modularity, small worldness or enhancement of the synchronization is a big success of this approach which gives confidence in the basic biological assumptions. On the contrary, the "random-wired" models
only impose synchronization  and are interested in the consequences of this requirement either on the network topology or on the link dynamics. For the latter, three main results have been reported: first a local synchronization process does give rise to a global synchronization, second during the adaptation toward synchronization, networks quickly exhibit community structures in which nodes of similar native frequencies form cliques. Then the synchronization of these cliques (and then their disappearance) is observed but on much longer time scales. Third,  the distribution of the native frequency and the distribution of the incoming weights turn out to be strongly correlated. The same observation is reported in \cite{Papadopoulos2017} which is "bio inspired". 

Now we review in more detail the assumptions used in the random-wired models \cite{Brede2008a, Brede2008b} :
\begin{enumerate}
\item First the network is assumed to be non-directed (ND). Although numerous neural network observations \cite{Song2005} have confirmed that bidirectional connections are more common than expected in a random network, this does not represent the totality of the connections anyway. Therefore, an excellent test of the validity of the modeling consists in not imposing the symmetry a priori and in checking that it then appears spontaneously during the adaptation phase of the network. 
\item The connection weight can only take 2 values (2V). The reason for this hypothesis is that it allows us to make the link with graph theory and to use the numerous tools then available to describe the network topology (smallness, clustering, modularity...). It turns out that the high precision of contemporary tract tracing can now resolve neural connections approximately equivalent to a single axonal projection and approximately a million times weaker than the strongest anatomical connections or white matter tracts \cite{Basset2017}. The existence of so many weak connections, reflected in the log normal distributions of connectivity weight \cite{Ercsey2013, Oh2014} casts strong doubt about the 2V assumption.
\item The maximum value that the weights can take is fixed at the beginning of the adaptation process. This constitutes a pre-established upper limit (PSUL) assumption. Note first that PSUL is not forced by 2V. Indeed one can very well imagine an adaptation process where the weights can only take 2 values, $W_{1}=0$ and $W_{2}\ne 0$, but where  $W_{2}$ is a variable which evolves over time. For example, in a pruning process, $W_{2}$ would increase while the number of non-zero connections would decrease. Note also that relaxing the PSUL assumption does not mean that the weights are going to diverge during the adaptation. More simply, setting an upper limit for the weights at the beginning of the process limits the trajectories and constrains the accessible configurations. 
\item The total mass of the connection weights is constant (CTM).  This is an extremely strong assumption
which even contradicts typical well known network mechanisms such as the Barab\`asi-Albert's \cite{Barabasi1999} model of preferential attachment or the episodes of synaptic connections pruning observed during learning processes \cite{Sakai2020}.
\end{enumerate}

In the next section we introduce an algorithm to compute synchronized networks of non-identical Kuramoto oscillators, with a total connection weight as small as possible, but where these 4 previous assumptions (ND, 2V, PSUL, CTM) are relaxed.
\section{Algorithm}
\label{Algorithm}
There are no strict and rigorous rules leading to the choice of the algorithm used. Rather, it is the result of a set of general considerations, analogies and heuristic arguments that we present below. Ultimately, the main rationale is that it effectively leads to synchronized solutions that are less expensive than the Kuramoto critical threshold.

\begin{enumerate}
\item We have deliberately chosen not to use a central control capable of accepting or rejecting a solution based on a global computation. The reason is that this kind of approach quickly becomes impractical with increasing $N$. On the contrary, we opted for a local, scalable and parallelizable approach. 

\item Following H.A. Simons' ideas in his famous paper "Architecture of the complexity", the nodes of the network were imposed to be unable to perform complicated mathematical computations (such as, for example, gradient computations or predictions). We just expect each oscillator to be able to estimate its degree of synchronization with each of its neighbors.

\item Each node can act locally to increase its synchronization with its neighbors. However, because of the lack of central control, the precise moments when each node decides to act are not synchronized. At each step of the adaptation process, only a small fraction of the nodes (we took $10 \% $ , randomly chosen) make an attempt to improve their synchronization.

\item Another consequence of the lack of central control: a node that has just tested a new local configuration but which ultimately does not adopt it, cannot force the rest of the network to return to its initial configuration. This would require too much effort in terms of storage and transport of information. The node that did the test must continue on its way. Optimization must be done on the fly.

\item A priori each node $i$ of the network can exercise control over its synchronization with its neighbors via the weight of the incoming $W_{ij}$ and/or outgoing $W_{ki}$ connections. We have chosen to intervene only on incoming connections. Our argument (far from being irrefutable) is the following: suppose that all the neighbors of $i$ have the same natural frequency. Then a slight increase  in incoming connections can easily lead to the synchronization of $i$ with its neighbors. On the other hand, synchronizing all the neighbors of $i$ with its natural frequency can require a strong increase in outgoing connections, not only because of the unfavorable ratio number ($1/N$) but also because the neighbors of $i$ are also subjected to the action of nodes which are not neighbors of $i$. Therefore the control of the synchronization through the incoming connections seems to be more efficient.

\item We have chosen not to impose any a priori structure on the connection network. Each node is connected with all the others but the weight of the connections evolves without constraint, can vanish or, on the contrary, grow indefinitely. This is a very expensive choice in terms of computating time but which is absolutely necessary to let the network freely choose its own topology. 

\end{enumerate}
Now we discuss the concrete implementation of the algorithm. 

Consider a system of $N$ non-identical Kuramoto's oscillators (eq.\ref{KuramotoEquation}), all to all connected. As in  \cite{Brede2008b} we introduce:
 \begin{equation}
 \begin{array}{c}
R(i)={{1}\over{T}}  \int_{t}^{t+T} \displaystyle{\sum_{j=1}^{N} \vert \partial_{t}\theta_{j}-\partial_{t}\theta_{i}\vert}dt \ge 0
\cr
R_{tot}=\displaystyle{\sum_{i=1}^N} R(i)
\end{array}
\end{equation}
as a measure of the synchronization of node $i$ with its neighbours. T is an arbitrary time interval associated with the periodicity with which the network adapts. Throughout the rest of our study $T$ is constant and fixed at $1$ (unit of time).
As we will be interested in the mass of connections, we also define
\begin{equation}
\begin{array}{c}
M_{in}(i)=\displaystyle{\sum_{j=1}^{N} W_{ij}}
\qquad 
M_{out}(i)=\displaystyle{\sum_{j=1}^{N} W_{ji}}
\cr
\displaystyle{\sum_{i}M_{in}(i)}=M_{tot}=\displaystyle{\sum_{i}M_{out}(i)}
\end{array}
\end{equation}
where $M_{in}(i)$ is the mass of the connections that point to the node $i$ while $M_{out}(i)$ is the mass of the connections emitted by $i$. Note that both $R(i)$, $M_{in}(i)$ and  $M_{out}(i)$ are local measurement at $i$.

At each rewiring step $n$, the adaptation process takes place in the following way:
\begin{enumerate}
\item For a given network configuration $W_{ij}^{n}$, a temporal evolution of (eq.\ref{KuramotoEquation}) between $t$ and $t+T$ is performed 
\item Same as the first step (between $t+T$ and $t+2T$), but now we also compute $R^{n}(i)$ and $M_{in}^{n+1}(i)$ for all nodes.
\item Next a fraction $p=0.1$ of the nodes are randomly selected. Let us call ${\cal S}$ this set.  Then a new network configuration $W_{ij}^{n+1}$ is generated where only the nodes $i_{sel}$ belonging to ${\cal S}$ are modified according to
\begin{equation}
W_{ij}^{n+1}=
\left\{
\begin{array}{lr}
W_{ij}^{n}  & i \notin {\cal S}
\cr
max \left( 0, W_{ij}^{n}+f \, rv(i,j) \right) & i \in {\cal S}
\end{array}
\right.
\end{equation}
where $rv()$ is a random variable uniformly distributed in $[-1,1]$, $f>0$ stands for the amplitude of the change and the $max()$ function ensures that the weights of the connections always remain positive. 
\item With the new configuration $W^{n+1}_{ij}$, a time evolution of (eq.\ref{KuramotoEquation}) between $t+2T$ and $t+3T$ is performed.
\item With the new configuration $W^{n+1}_{ij}$, time evolution of (eq.\ref{KuramotoEquation}) between $t+3T$ and $t+4T$ with measurement of $R^{n+1}(i_{sel})$ and $M_{in}^{n+1}(i_{sel})$ only for the nodes $i_{sel}$ in ${\cal S}$.
\item For each node $i_{sel} \in {\cal S}$, the local network modification ($W^{n}(i_{sel}j) \longrightarrow W^{n+1}(i_{sel}j) $) is accepted if and only if one of the following two conditions is fulfilled:
\begin{equation}
\begin{array}{l}
(A) \, \left\{
\begin{array}{l}
R^{n}(i_{sel})=R^{n+1}(i_{sel})=0 
\cr
M_{in}^{n+1}(i_{sel})<M_{in}^{n}(i_{sel})
\end{array}
\right.
\cr
\cr
(B) \, \left\{
\begin{array}{l}
R^{n}(i_{sel}) \ne 0 
\cr
R^{n}(i_{sel})>R^{n+1}(i_{sel})
\end{array}
\right.
\end{array}
\label{decision}
\end{equation}
Condition (B) ensures that a stable configuration is sought as a priority, and that only after this condition has been met that economic considerations (A) is taken into account. When none of theses conditions are fulfiled, the local network modification $W^{n+1}(i_{sel}j) $ is rejected and the algorithm restarts from the beginning with $W^{n}(i_{sel}j)$.
\end{enumerate}
Some remarks are in order: i) A forth order Runge Kutta scheme with an adaptative time increment  is used. ii) Adaptation is done on the fly and the intermediates stages 1 and 4 are expected to allow the oscillators to converge to an asymptotic state. iii) In the numerical simulation, the condition $R=0$  is substituted with the  more realistic one $R \le R_{thres}$. iv) As it is, the possible decrease of $R(i_{sel})$ forced by (B) is not necessarily accompanied by a decrease (or at least stationarity) of $R(j), j \notin {\cal S}$. Therefore, convergence towards a global synchronized network is far from being obvious. v) Instead of eq.(\ref{decision}), a more classic approach would have been to minimize $C(i_{sel})=R(i_{sel})+M_{in}(i_{sel}) / \delta$, 
where  $\delta$ is a characteristic positive mass which has to be suitably chosen. As $R(i_{sel})$ decreases and $M_{in}(i_{sel}) / \delta$ increases with $M_{in}(i_{sel})$,  then for large enough value of $\delta$ there exists $M^{min}_{in}$ for which $C(i_{sel})$ displays a minimum. We have performed numerical simulations with this choice and observed that it gives rise to clusters of synchronization with a typical $M^{min}_{in}$ mass. Global synchronization is reached only for large value of $\delta$. In a way, our choice (eq.(\ref{decision})) corresponds to the limit $\delta \longrightarrow \infty$.
\begin{figure}
\resizebox{0.40\textwidth}{!}{
\includegraphics[]{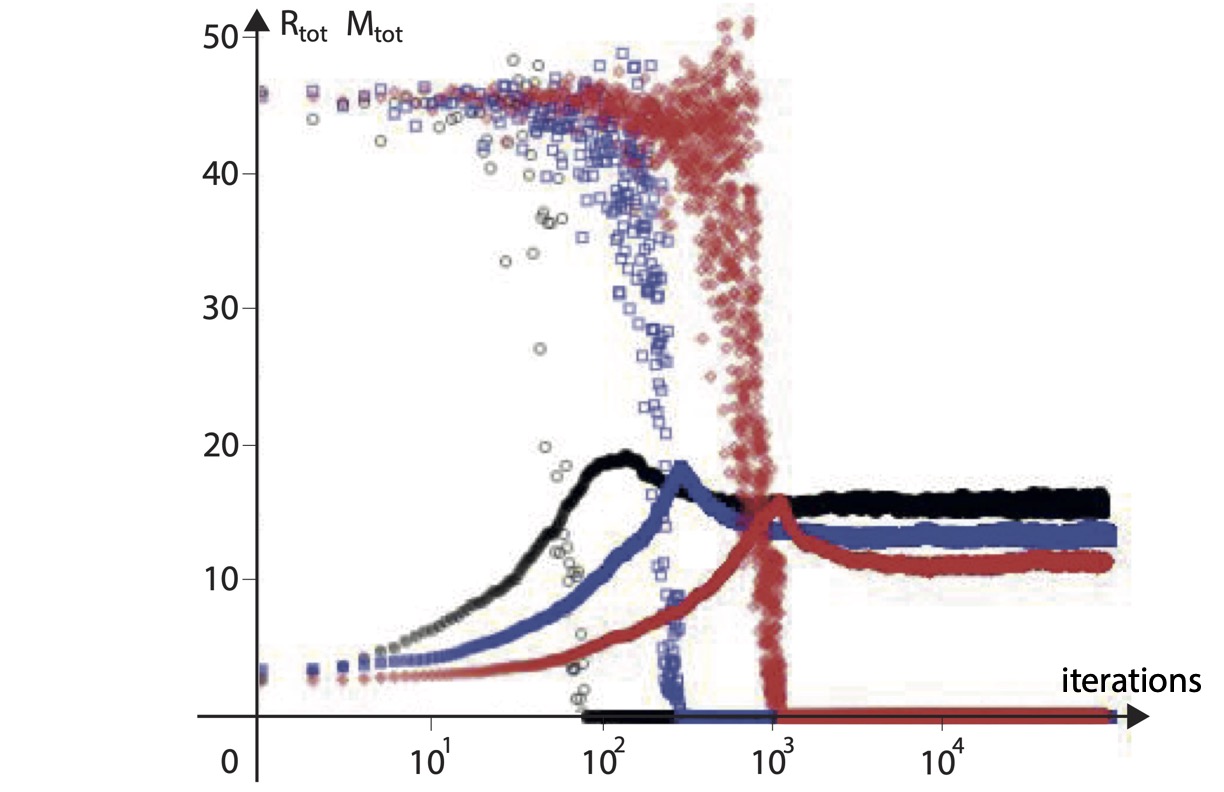}
}
\caption{$N=20$. The 3 top curves (with hollow symbols) display $R_{tot}$ (in linear scale) versus the number of adaption step (in log scale) for $f=0.2$ (black circles), $0.1$ (blue squarres) and $0.05$ (red circles). The 3 bottom curves (with solid symbols and the same color code) show the corresponding evolution of  $M_{tot}$.}
\label{fig1}
\end{figure}

\begin{figure}
\resizebox{0.40\textwidth}{!}{
\includegraphics[]{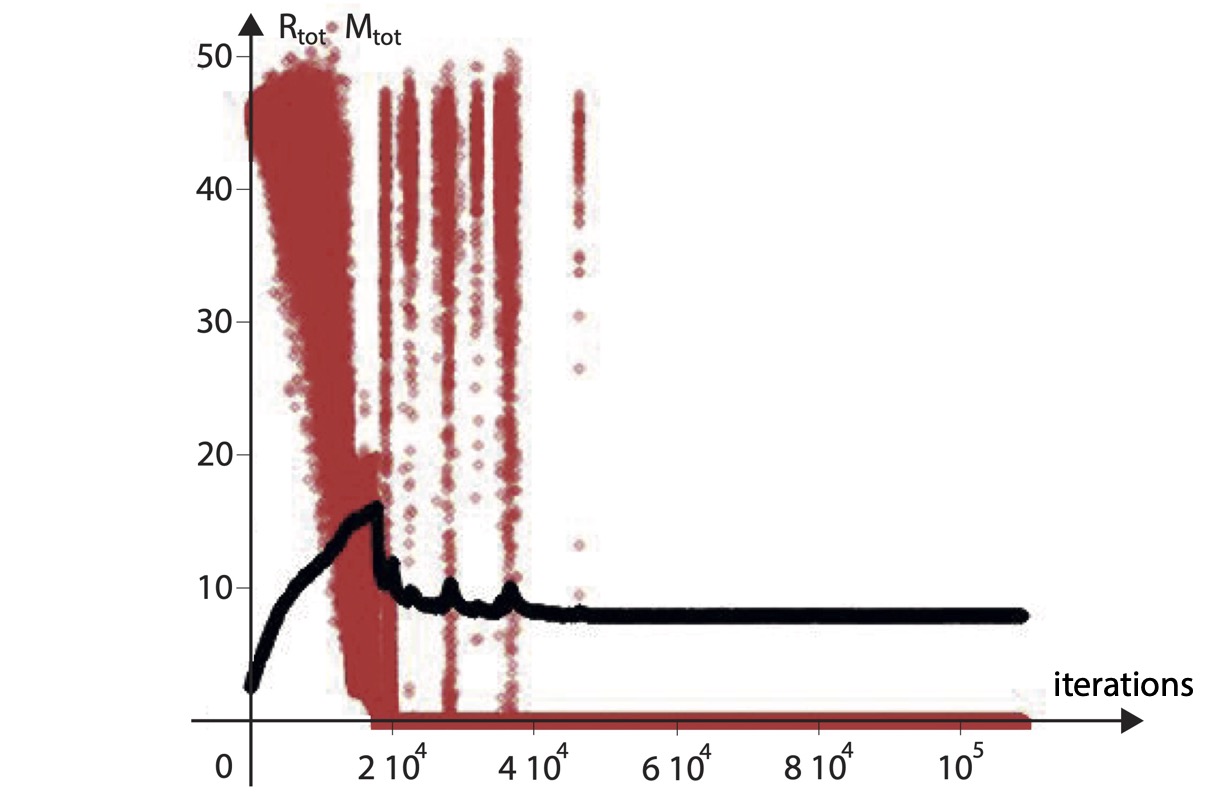}
}
\caption{$N=20$. Plots of $R_{tot}$ (red) and  $M_{tot}$ (black) versus the number of adaptation steps for $f=0.0125$. Note that $R_{tot}$ does not decrease continuously, but rather presents episodes of strong fluctuation. Indeed, when the network is synchronized ($R_{tot}=0$), the continuous search for a lower total mass may suddenly drive the network out of synchronization. This leads to a rapid increase in total mass until a synchronized regime is reached again.}
\label{fig2}
\end{figure}

\begin{figure}
\resizebox{0.40\textwidth}{!}{
\includegraphics[]{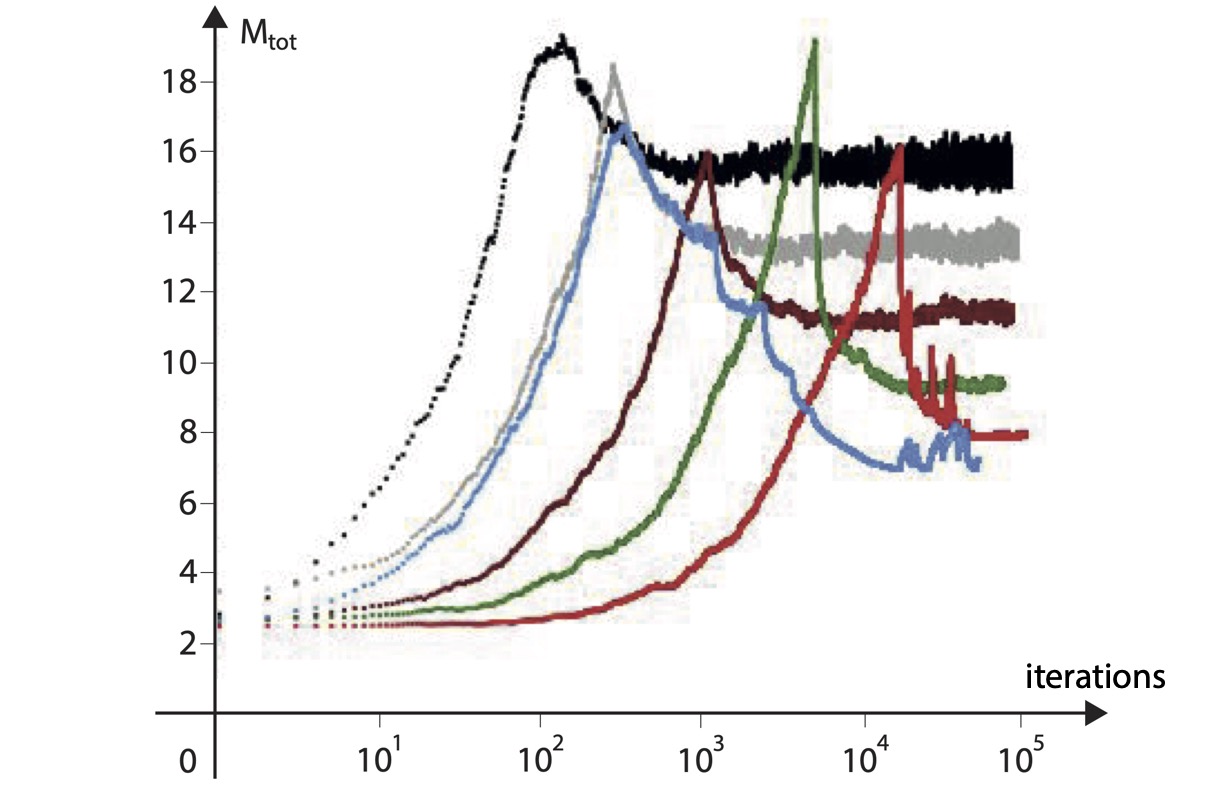}
}
\caption{$N=20$. Plots of $M_{tot}$ versus the number of adaptation steps (log scale) for several values of $f$ ($f=0.2$ black, $0.1$ gray, $0.05$ maroon, $0.025$ green and $0.0125$ red). For the curve in blue, $f$ has successively been fixed to $0.1$, $0.05$, $0.025$, $0.0125$, $0.00625$, $0.00425$, $0.00325$ and $0.00225$ during the same adaptation process. }
\label{fig3}
\end{figure}

The plots in fig.\ref{fig1}-\ref{fig3} show the typical time evolution of $R_{tot}$ and $M_{tot}$ during the adaptation process of a network of $N=20$ Kuramoto's  oscillators. The various curves correspond to the same natural frequency sample and the same random distribution of initial connection weights and initial phase $\theta_{i}(0)$, only $f$ is changed.
The first and most important observation is that the algorithm effectively converges to a fully synchronized network with an almost vanishing  $R_{tot}$ (fig.\ref{fig1}). Because the algorithm is continuously testing new configurations, $R_{tot}$ is not exactly equal to zero but fluctuates around this value. However, we check each time that imposing f=0 from such a configuration leads systematically and accurately to  $R_{tot}=0$. The second observation is that the final $M_{tot}$, and therefore the adapted network, depends on $f$: the smaller $f$ the smaller the final mass.  Third, the curve of the total mass versus time shows a maximum : as the initial connection weights are too small to  sustain synchronization, there is first an increase in the mass of connections until a synchronized regime is reached followed by a phase of adjustment of the weight of the connections. Note that this temporal evolution of the total mass is in strong disagreement with the CTM assumption in \cite{Brede2008a,Brede2008b}. Finally by continuing to reduce the weight of the connections, it can happen that the network ends up not being synchronized anymore. In this case, an avalanche dynamics is observed (fig.\ref{fig2}), characterized by slow mass increase phases, alternating with periods of synchronization and rapid mass decay. In this case the algorithm detects and saves the network configuration corresponding to $R_{tot}=0$ and the smallest total mass. All of these observations suggest that faster and more accurate convergence can be achieved by varying $f$ during the adaptation process (as in fig.\ref{fig3} ). Such a strategy will be used in the rest of the study.

\begin{figure}
\resizebox{0.40\textwidth}{!}{
\includegraphics[]{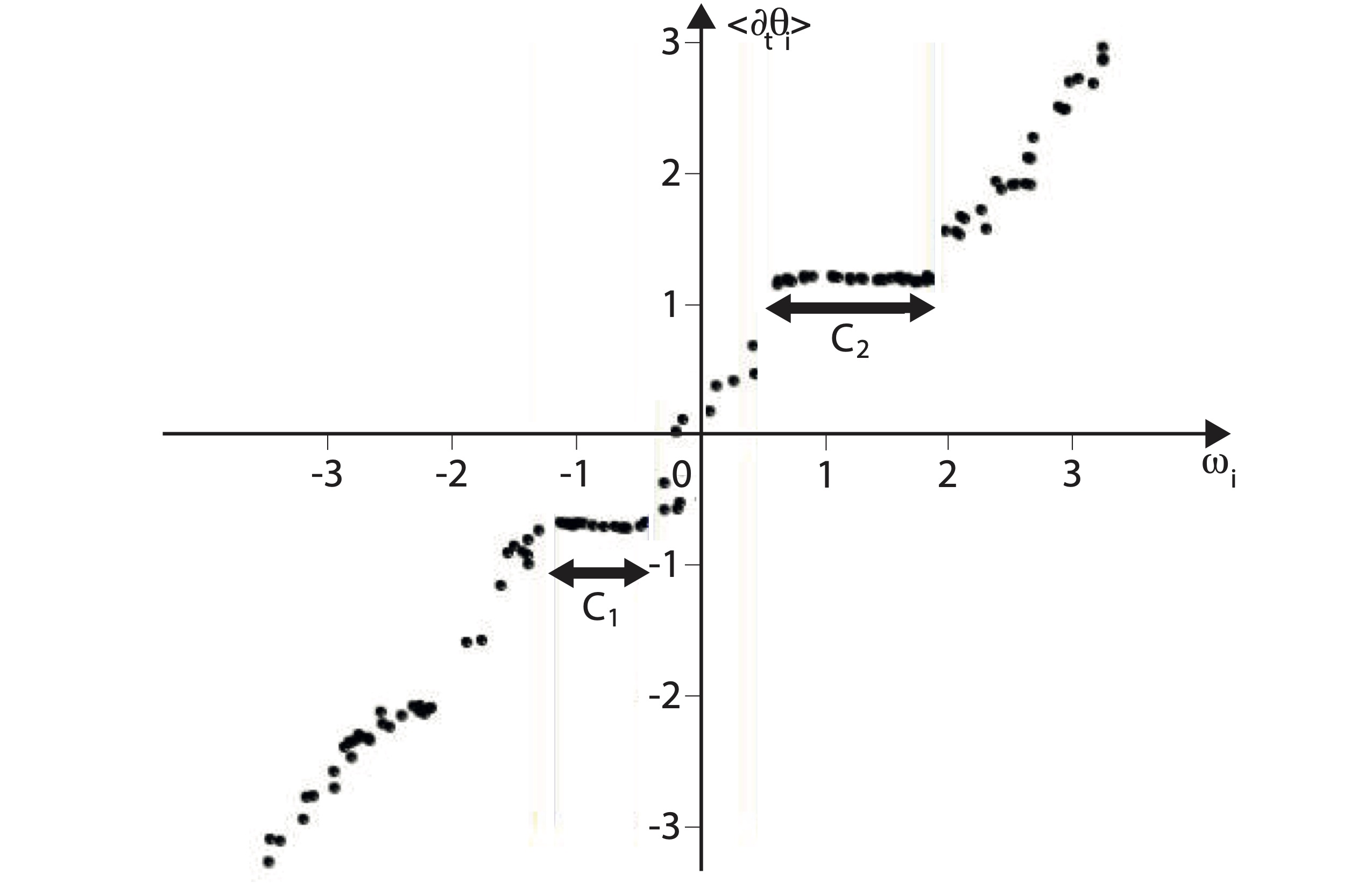}
}
\caption{$N=120$. Plot of the average frequencies \hbox{$<\partial_{t} \theta_{i} >_{time}$} versus their natural frequency $\omega_{i}$ after 200 adaptation steps. ${\cal C}_{1}$ designates one group of synchronized oscillators, ${\cal C}_{2}$ another.}
\label{fig4}
\end{figure}

\begin{figure}
\resizebox{0.40\textwidth}{!}{
\includegraphics[]{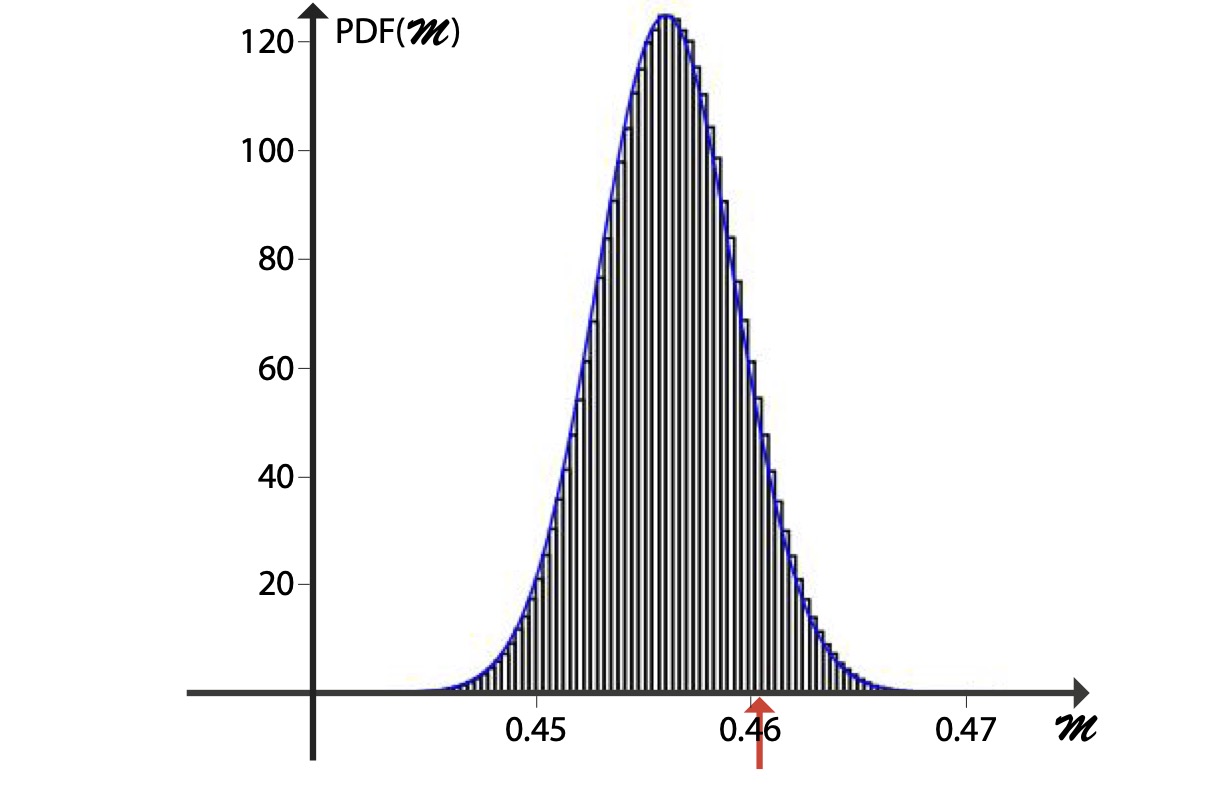}
}
\caption{$N=120$. The black histogram is the PDF of ${\cal M}(\pi)$ built from $10^6$ achievements for the partition ${\cal P}=\left\{ {\cal C}_{1}, {\cal C}_{2}, {\cal C}_{3} \right\} $. The blue line is a gaussian fit, and the vertical red arrow points to the measured value ${\cal M}$ (eq.\ref{Modularite}). }
\label{fig5}
\end{figure}

During the adaptation process, transient structures appear, related to the fast synchronization of disjoint groups of oscillators whose natural frequencies are close (fig.\ref{fig4} and fig.\ref{fig5}). Fig.\ref{fig4} displays the time average frequency $<\partial_{t} \theta_{i} >={{1}\over{T}} \int_{t}^{t+T} \partial_{t}\theta_{i} dt$ of $N=120$ Kuramoto's oscillators versus their natural frequency $\omega_{i}$, some time after the beginning of the process. The plot clearly shows the existence of two disjoint groups of oscillators  (${\cal C}_{1}$ and ${\cal C}_{2}$) whose members have almost the same effective frequency. We call ${\cal C}_{3}$ the complementary set of ${\cal C}_{1} \cup {\cal C}_{2}$. To check if the partition ${\cal P}=\left\{ {\cal C}_{1}, {\cal C}_{2}, {\cal C}_{3} \right\} $ actually forms a community, we first compute
\begin{equation}
{\cal M}=\displaystyle{ {{1}\over{M_{tot}}}  \sum_{{\cal C}_{k} \in {\cal P}} \sum_{  (i,j) \in {\cal C}_{k}^2   } W_{ij}} 
\label{Modularite}
\end{equation}
which measures the relative importance of intra-community connections. For example, when there are no connections between the ${\cal C}_{i}$, ${\cal M}$ is maximum and equal to $1.0$. However, more than its numerical value in itself, what is important is to compare the result to the value ${\cal M}_{rand}$ it would have for any partition of 3 distincts sets $\left\{Q_{1}, Q_{2}, Q_{3}\right\}$ with cardinals: $\#(Q_{i})=\#({\cal C}_{i}),i=1..3$ but constituted at random. Note that, contrary to the graph theory case where this latter value can be analytically derived, here the distribution $W_{ij}$ is only numerically known. This is a difficulty that we will face several times throughout this study such it's worthwhile, at least for this first time, to proceed in a detailed way: First we randomly generate a permutation $\pi$ of the set $\left\{1,2...N\right\}$  and apply it on $\left\{ {\cal C}_{1}, {\cal C}_{2}, {\cal C}_{3} \right\}$ to create a new partition $\left\{ {\cal Q}_{1}, {\cal Q}_{2}, {\cal Q}_{3} \right\}$ with the same number of elements. Then we compute ${\cal M}(\pi)$ for a large number of random permutations $\pi$ and derive a well-defined mean value ${\cal M}_{rand}$ and standard deviation $\sigma$. Fig.\ref{fig5} shows that the resulting probability density function (PDF) of ${\cal M}(\pi)$ obtained after $6 \, 10^6$ random permutations is very close to a Gaussian curve. The convergence is very fast and we do check that the result is unchanged for $5 \, 10^5$, $10^6$ and $6 \, 10^6$ random draws. Numerically we find ${\cal M} \simeq 0.46$ which is only about $\simeq 1.36$ standard deviations from the mean value and does not exclude a result only due to chance. Therefore the existence of a community is not a sure result.

\section{Results}
\label{Results}
\subsection{Economy}

\begin{figure}
\resizebox{0.40\textwidth}{!}{
\includegraphics[]{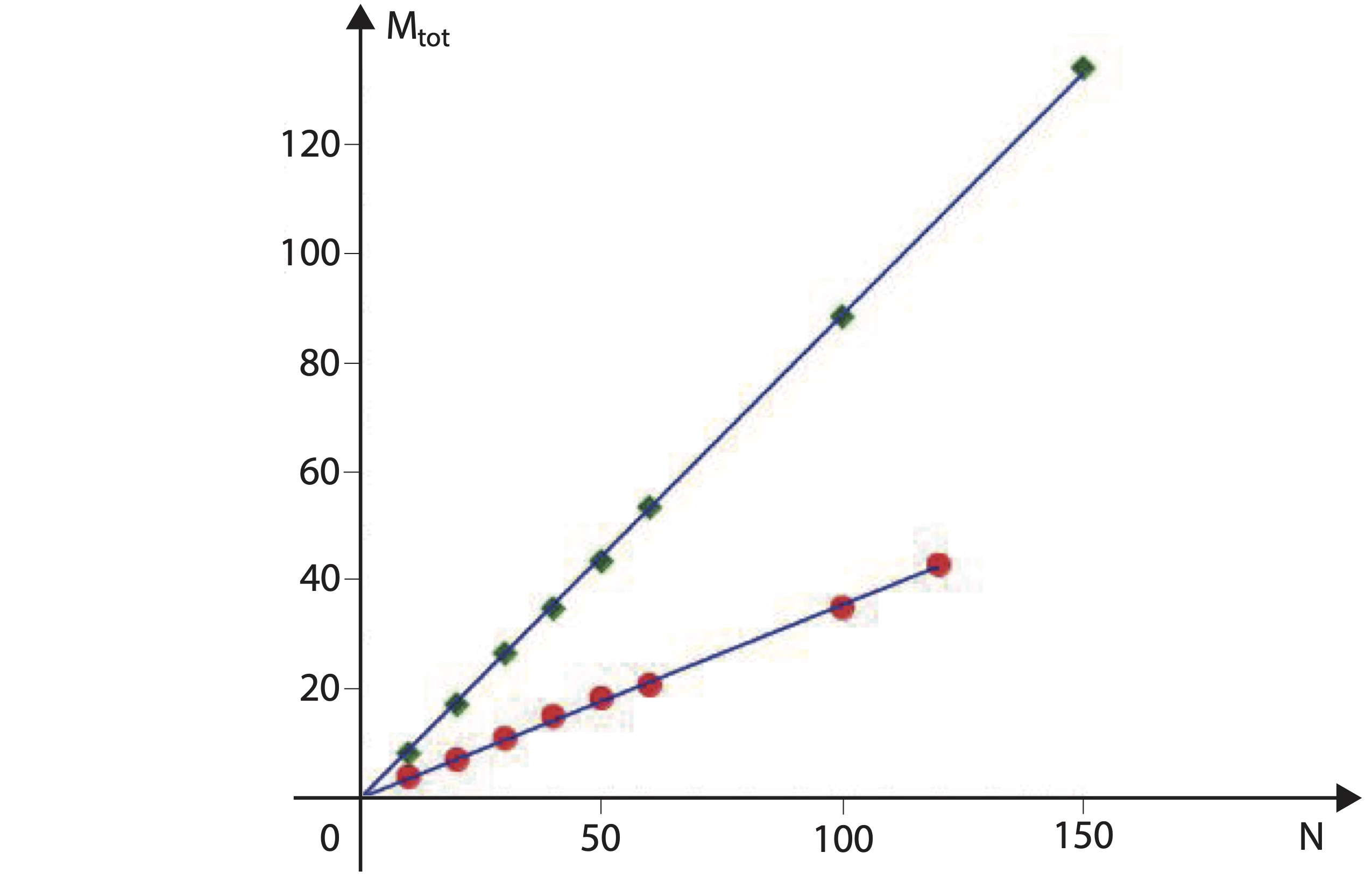}
}
\caption{Total mass of the connection weights (${\cal M}_{tot}$) versus the network size $N$. The green diamonds correspond to the case of an homogeneously connected Kuramoto's network of non-identical oscillators at the synchronization threshold, the red solid circles to the adapted network. In both cases, the same sample of natural frequencies is used. The continuous lines are linear fits and the ratio between the slopes of these straight lines is about $0.41 \pm 0.03$.  }
\label{fig6}
\end{figure}

The condition (A) in (eq.\ref{decision}) decreases the mass of the incoming connections as soon as it possible. Therefore we expect and we do observe (fig.\ref{fig6}) that the total mass of the network after adaptation is well below ($\simeq 40 \% $) that of the homogeneous Kuramoto network with the same sample of natural frequencies, at the threshold of the synchronization transition. Note that: i) The Kuramoto's threshold has been analytical derived for an infinite network. But as the networks we consider are rather small ($N \le 120$), the precise value of Kuramoto threshold has to be numerically determined. ii) In order to minimize statistical fluctuations on natural frequency samples, we proceed as follows: first we generate a sequence $\Omega'=\{\omega'_{i},i=1..N\}$ of $N$ frequencies between $-1$ and $+1$ with a uniform random generator. Then the sample average frequence ($<\omega'> \simeq 0$) and standard deviation ($\sigma' \simeq 0.57$) are computed. Finally a new set of natural frequencies is derived through $\omega_{i}=({{2}\over{\sigma'}})\left(\omega_{i}'-<\omega'>\right)$. In this way, whatever $N$ and the random draw, the resulting natural frequency samples all have exactly a mean value of zero and a standard deviation of 2. iii) For $N=100$, we have tested 10 independent samples of natural frequency and measure that the final total mass fluctuations are smaller than the corresponding symbol size in fig.\ref{fig6}.

\subsection{Strong correlation between the natural frequency distribution and the weight distribution of both incoming and outgoing connections}

\begin{figure}
\resizebox{0.40\textwidth}{!}{
\includegraphics[]{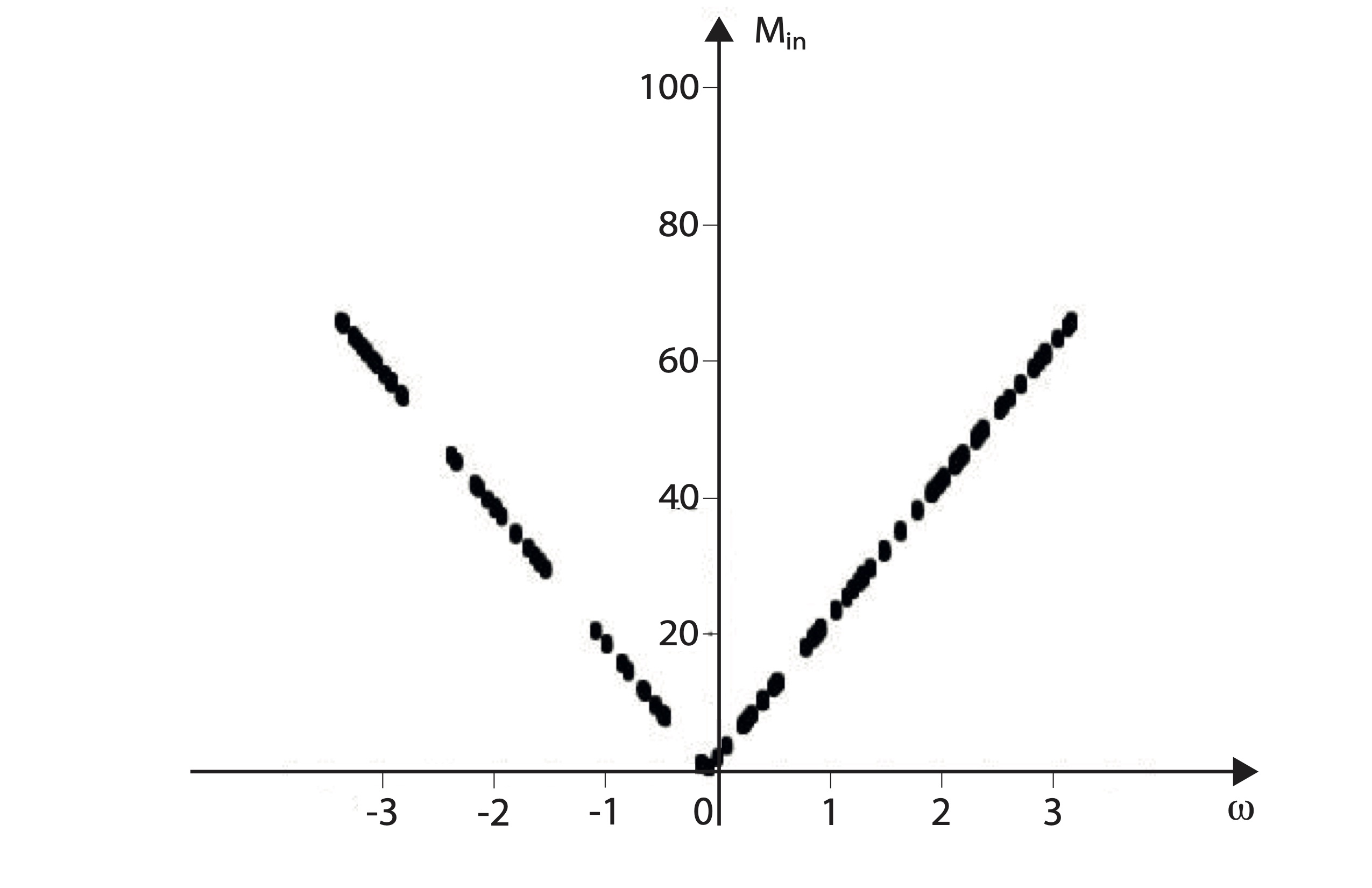}
}
\caption{ $N=100$. Plot of the incoming connection mass ${\cal M}_{in}(i)$ versus $\omega_{i}$. The figure cumulates the results of 10 simulations associated with the same sample of natural frequencies, but with different initial conditions and realizations. The natural frequencies are uniformly distributed, with a mean value exactly zero and a standard deviation precisely equal to 2.}
\label{fig7}
\end{figure}

\begin{figure}
\resizebox{0.40\textwidth}{!}{
\includegraphics[]{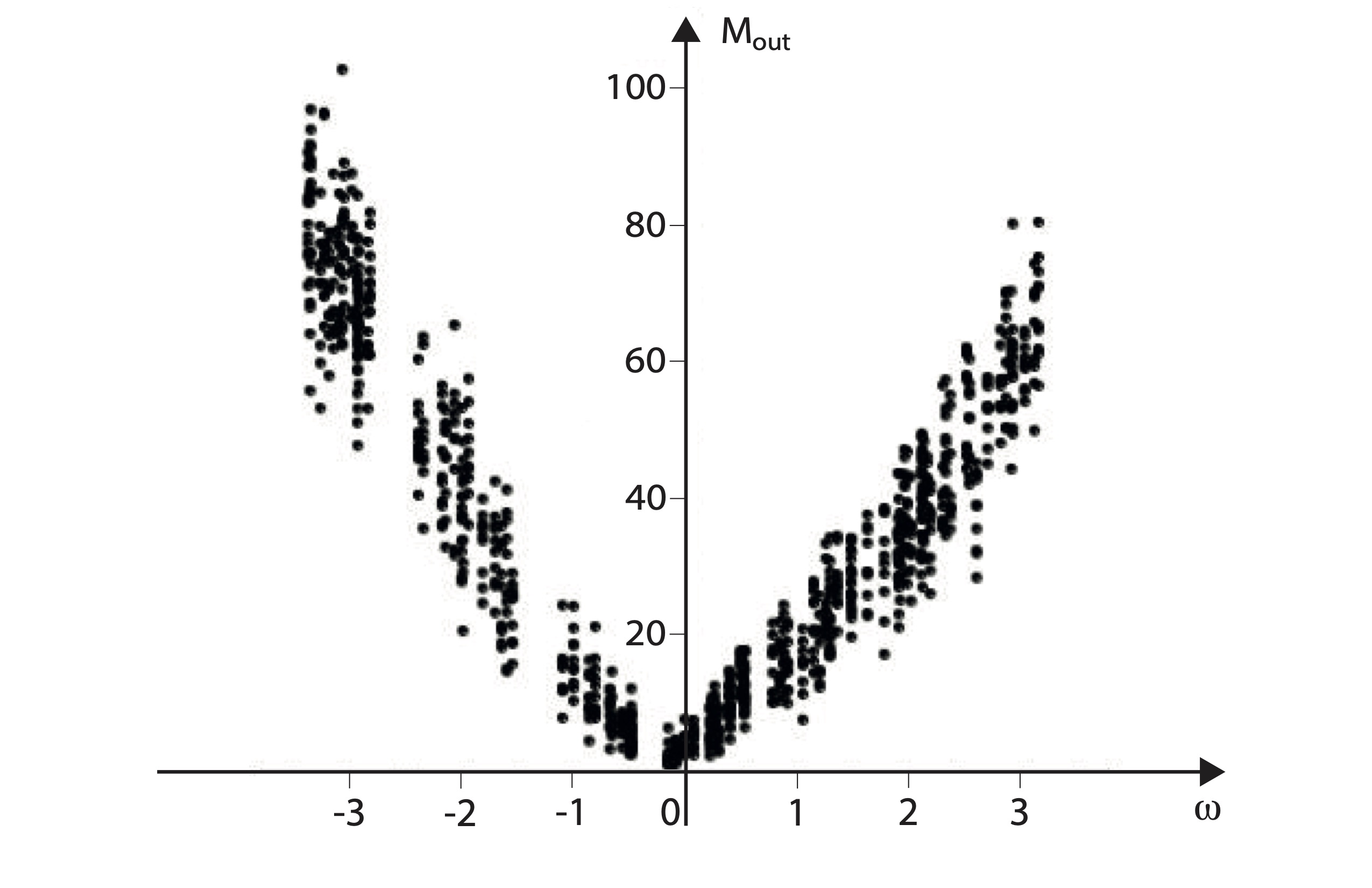}
}
\caption{Same as fig.\ref{fig7}, but now the plot displays the outgoing connection mass ${\cal M}_{out}(i)$ versus $\omega_{i}$. Note the vertical scale is the same as in fig.\ref{fig7} in order to facilitate the comparison of the interval range.}
\label{fig8}
\end{figure}

\begin{figure}
\resizebox{0.50\textwidth}{!}{
\includegraphics[]{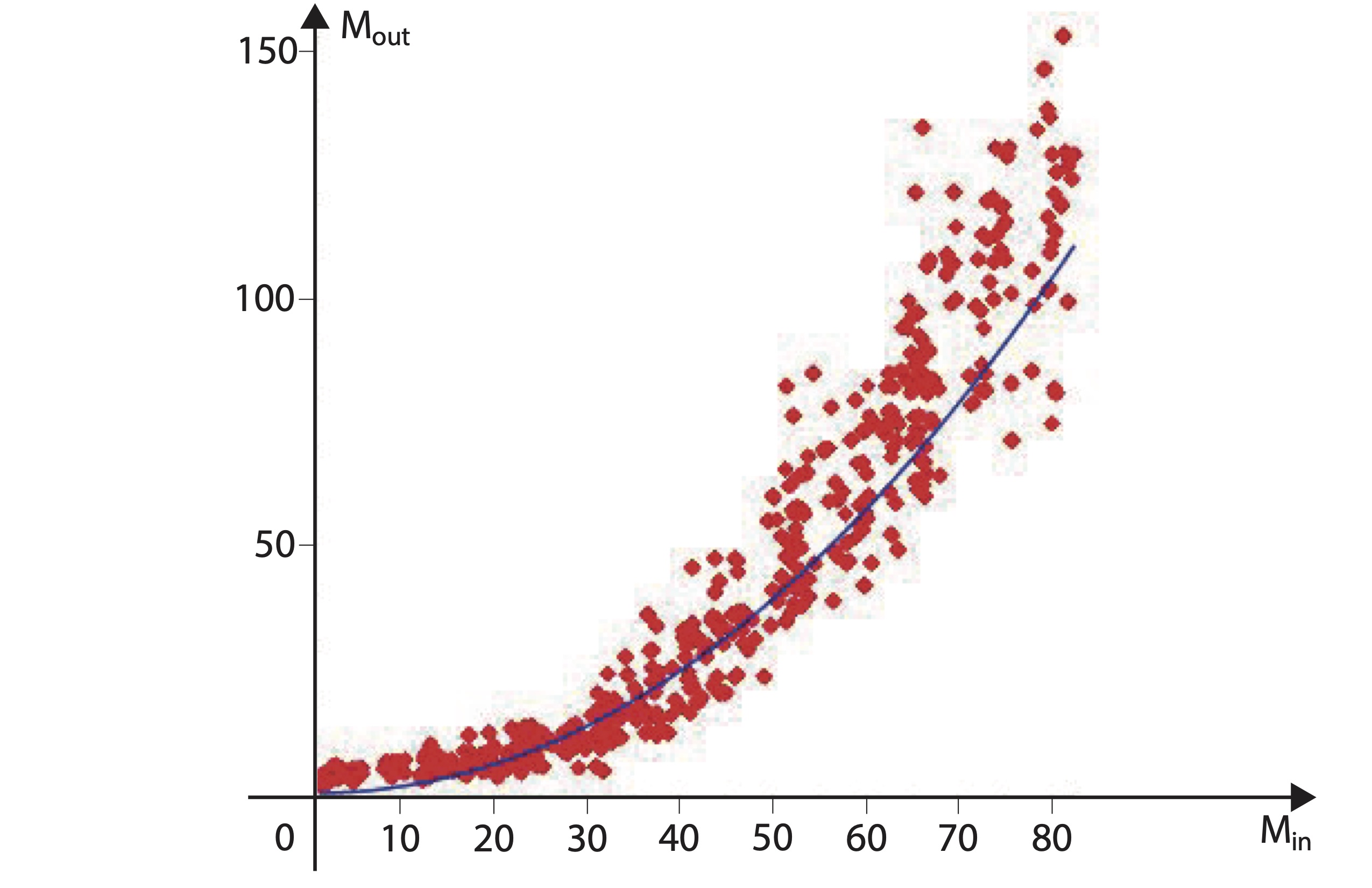}
\includegraphics[]{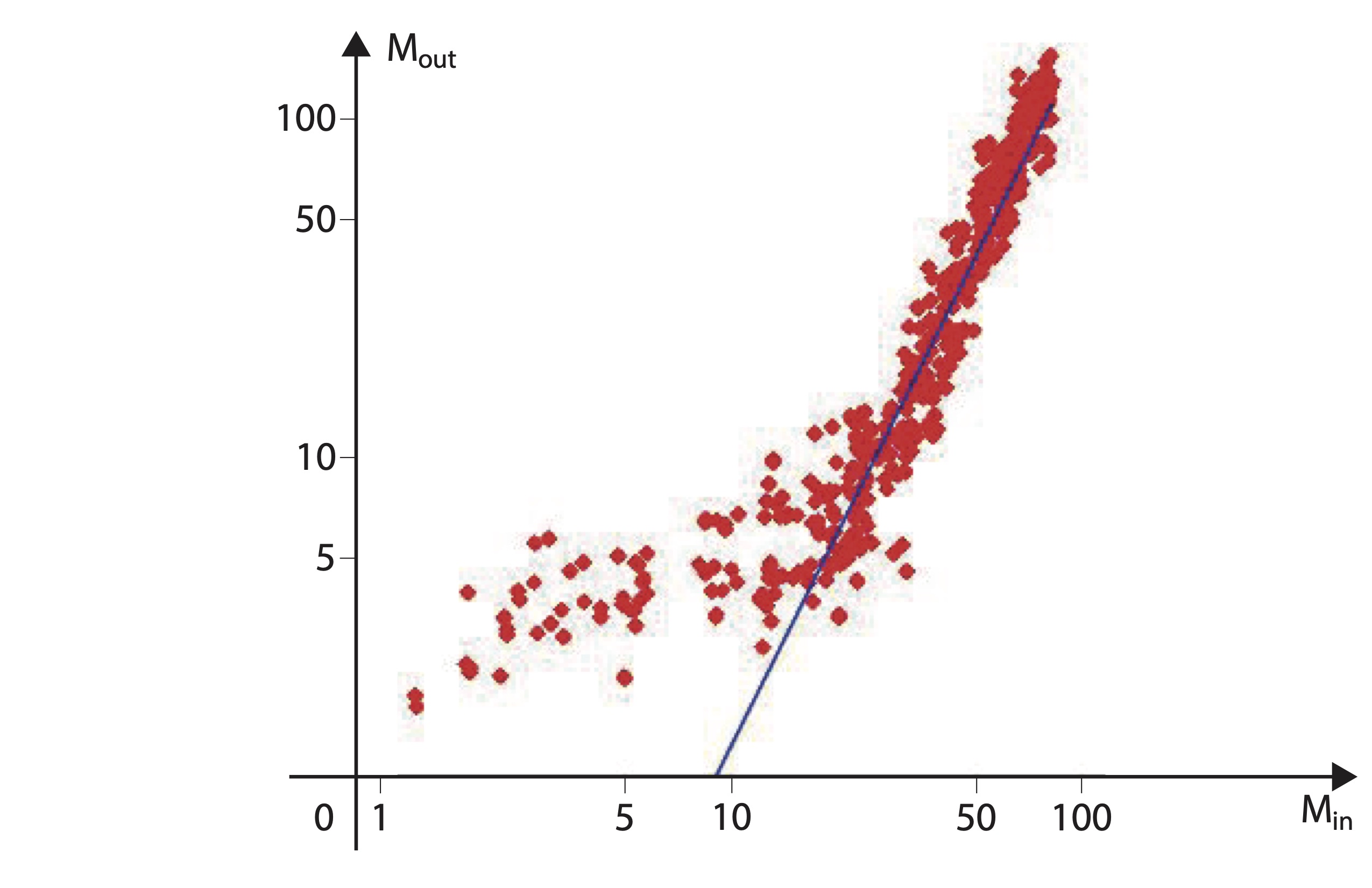}
}
\caption{Same as fig.\ref{fig7} but for $N=120$. The plots display the same outgoing connection mass ${\cal M}_{out}(i)$ versus the incoming connection mass ${\cal M}_{in}(i)$, either in a lin-lin or log-log scales. The continuous black line is a power law fit with an exponent  $\simeq 2$.}
\label{fig9}
\end{figure}

A ratio of 40 \% between the total mass of the adapted network and that of the Kuramoto network (fig.\ref{fig6}) suggests an excellent adaptation of the network geometry to the inhomogeneity of the natural frequency distribution. As already reported in \cite{Brede2008a,Brede2008b,Papadopoulos2017}, we also observe the appearance of the marked v-shaped curves characterizing the plots of ${\cal M}_{in}(i)$ vs. their natural frequency $\omega_{i}$ (fig.\ref{fig7}).  Never reported before, fig.\ref{fig8} and fig.\ref{fig9} show the existence of a strong parabolic relationship between outgoing weights ${\cal M}_{out}(i)$ and natural frequencies $\omega_{i}$. Note this dependence is at least as important as that of ${\cal M}_{in}(i)$ because the global variation intervals of ${\cal M}_{in}(i)$ and ${\cal M}_{out}(i)$ with natural frequencies are almost identical. 

\begin{figure}
\resizebox{0.35\textwidth}{!}{
\includegraphics[]{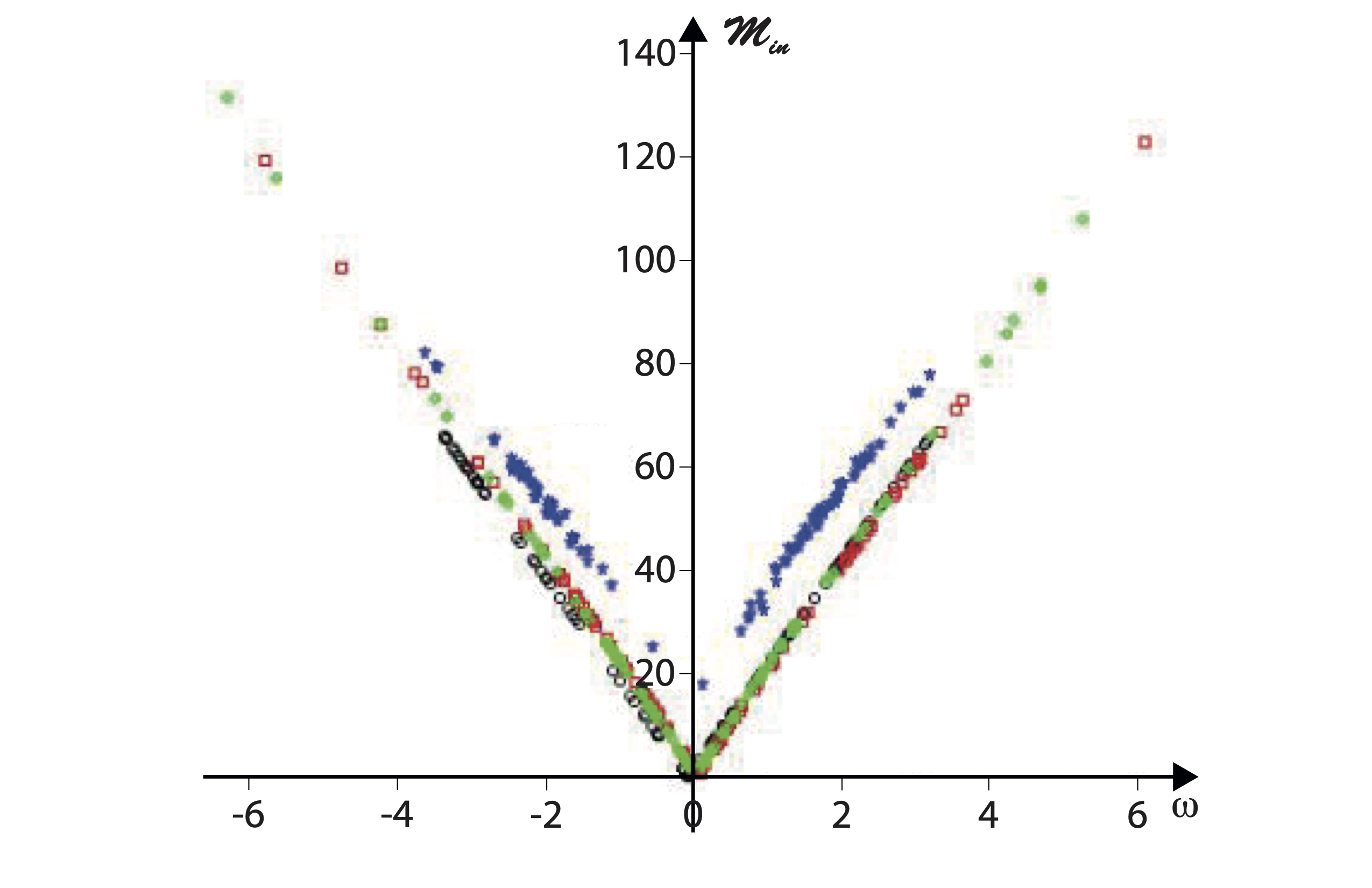}
}
\caption{$N=100$. The plot displays the incoming connection mass ${\cal M}_{in}(i)$ versus the natural frequency  $\omega_{i}$ for respectively a uniform (black circles), gaussian (red squares), cauchy (green diamonds) and bimodal (blue asterisks) distributions of natural frequencies.}
\label{fig10}
\end{figure}

\begin{figure}
\resizebox{0.35\textwidth}{!}{
\includegraphics[]{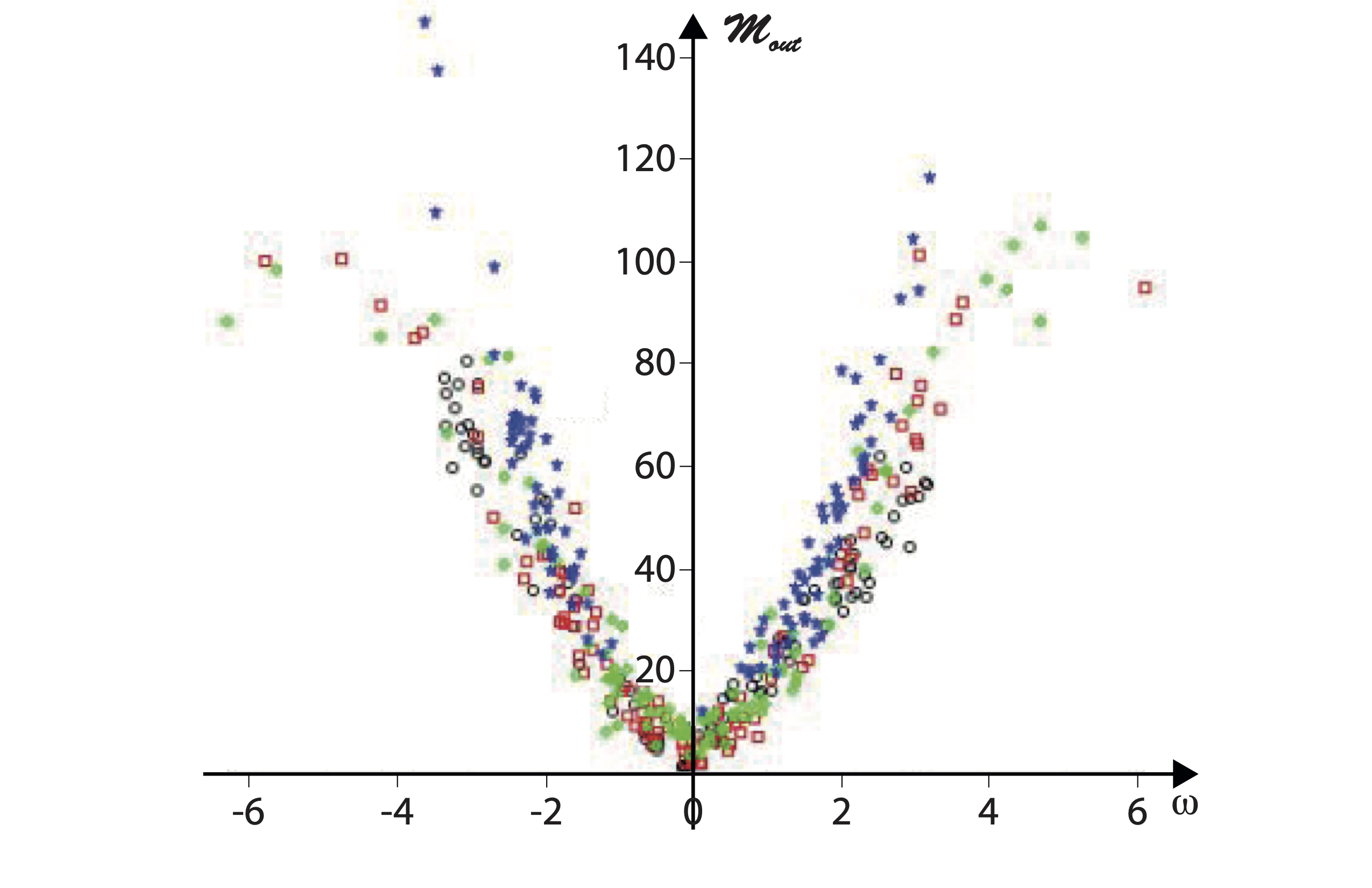}
}
\caption{Same as fig.\ref{fig10} but for the outgoing connection mass ${\cal M}_{out}(i)$.}
\label{fig11}
\end{figure}

The v-shaped curve has been interpreted \cite{Brede2008a,Brede2008b,Papadopoulos2017} as a clear tendency of the network to associate primarily below-average frequencies with above-average frequencies. The main idea is that the synchronization of 2 oscillators with respective frequencies $\omega_{1}$ and $\omega_{2}$ requires a total connection mass proportional to $\vert \omega_{2}-\omega_{1}\vert$. Once synchronized, this system oscillates with an intermediate frequency ${{W_{12} \omega_{2}+W_{21} \omega_{1}}\over{W_{12}+W_{21}}}$. In terms of connection weight, it is very costly to synchronize two oscillators with opposite frequencies, but once done, we end up with pairs that have almost identical vanishing frequencies and therefore are easily synchronizable. However, this interpretation does not explain the piecewise linear form of the ${\cal M}_{in}(i)$ versus $\omega$ curve, nor does it explain the parabolic dependence of ${\cal M}_{out}(i)$. We have verified that these results persist regardless of the shape (uniform, gaussian, cauchy, bimodal) of the natural frequency distribution (fig.\ref{fig10} and fig.\ref{fig11}).

\subsection{Rich community structure}
The oscillators with the natural frequencies furthest from the average value are those with the strongest network of both incoming and outgoing connections (fig.\ref{fig10} and fig.\ref{fig11}). But are these connections made between oscillators of the same type (assortativity) or on the contrary between oscillators of different types (disassortativity) \cite{Colizza2006}?

\begin{figure}
\resizebox{0.35\textwidth}{!}{
\includegraphics[]{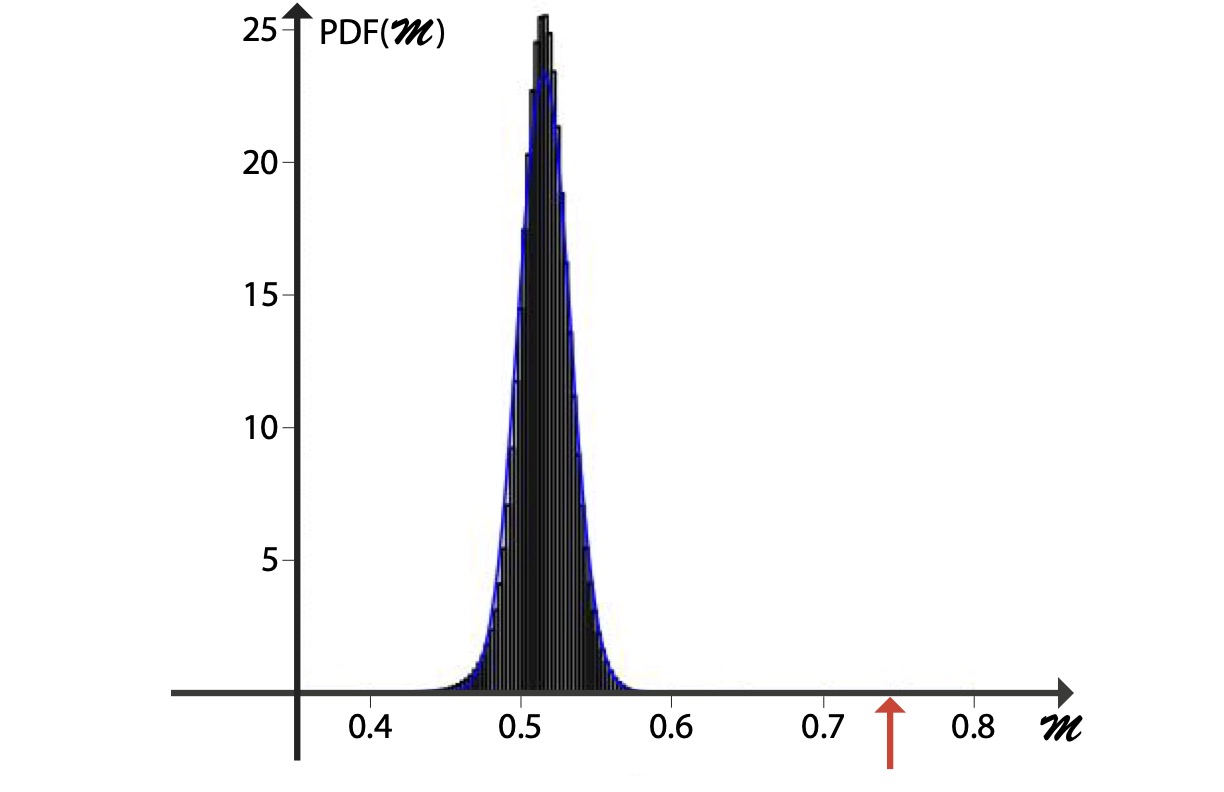}
}
\caption{$N=100$. The black histogram is the PDF of ${\cal M}(\pi)$  for the partition $\left\{ \Omega_{-}, \Omega_{+} \right\} $. It has been obtained after $6 \, 10^{6}$ random generations of permutation, which, compared to the gigantic number of permutations of $N \, (N-1)$ elements, is completely negligible. Nevertheless the convergence is very fast and we do check that the results for $5 \,10^4$, $10^5$, $10^6$ and $6 \, 10^6$ are strictly identical. The blue continuous line is a Gaussian fit with the same mean value and standard distribution as the measured PDF. The vertical red arrow points to the measured value ${\cal M}$ (eq.\ref{Modularite}): It is more than 13.4 standard deviations from the mean value.}
\label{fig12}
\end{figure}

To answer this question, we define $\Omega_{-}$ as the set of oscillators whose natural frequency belongs to $\left[ -\omega_{th},+\omega_{th}\right]$, and $\Omega_{+}$ as its complementary set (we arbitrary chose $\omega_{th}= 0.41 \, \omega_{max}$ where $\omega_{max}=sup \left\{ \vert \omega_{i} \vert \right\}$). To determine whether the partition $ \left\{ \Omega_{-},\Omega_{+} \right\} $ forms a community, we proceed in exactly the same way as when we tested whether the synchronized structures ${\cal C}_{1}$ and ${\cal C}_{2}$ in fig.\ref{fig4} formed a community. We compute the proportion of intracommunauty connection as in (eq.\ref{Modularite}) and compare the result with the value it would have for any  partition $ \left\{ \pi(\Omega_{-}),\pi(\Omega_{+}) \right\}$ where $\pi$ is a random permutation.  The corresponding computations are displayed in fig.\ref{fig12}. The measured value deviates from the mean value by more than $13.4$ standard deviations, which definitively rules out an observation due to chance alone.The result is therefore indisputable: the partition does form a community \cite{Colizza2006}. 

We have also considered the 3-sets partition $\left\{\Omega_{1},\Omega_{2},\Omega_{3}\right\}$ with
\begin{equation}
\begin{array}{llll}
\Omega_{1}=\{ i \in [1,N] ,  & \vert {{\omega_{i}}\over{\omega_{max}}} \vert &< 0.41 &\}
\cr
\Omega_{2}=\{ i \in [1,N] ,  0.41 \le & \vert  {{\omega_{i}}\over{\omega_{max}}} \vert &< 0.56  &\}
\cr
\Omega_{3}=\{ i \in [1,N] ,  0.56 \le &\vert  {{\omega_{i}}\over{\omega_{max}}} \vert  & &\}
\end{array}
\end{equation}
computed the proportion of intracommunauty connections  (eq.\ref{Modularite}) and found it deviates from the mean value by more than $9.4$ standard deviations. The adapted network definitely possesses a rich club structure where the most connected oscillators are preferentially connected to each other.

\subsection{Highly Symmetrical}
\begin{figure}
\resizebox{0.35\textwidth}{!}{
\includegraphics[]{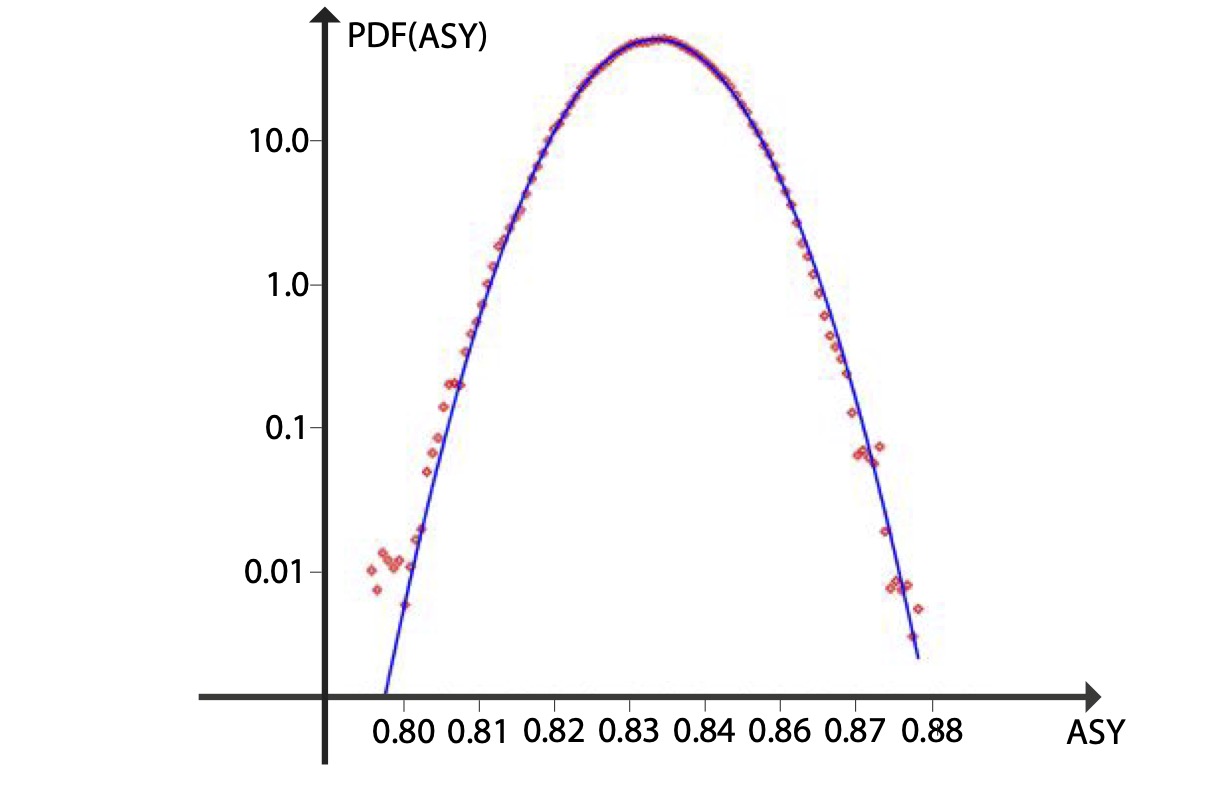}
}
\caption{$N=100$. PDF of $ASY(\pi)$  obtained after $10^{4}$(blue circles), $10^6$ (green box) and $10^7$ (red diamond) random generations of permutation $\pi$. The vertical scale is logarithmic. The measured $ASY$ value of the adapted network is $\simeq 0.48$ (not shown because well outside the frame), i.e. more than $44$ standard deviation below the mean value!}
\label{fig13}
\end{figure}

In order to quantify the degree of symmetry of the adapted network, we introduce
\begin{equation}
ASY={{1}\over{{\cal M}_{tot}}} \displaystyle{ \sum_{i,j} \vert W_{ij}-W_{ji} \vert}
\label{ASYDefinition}
\end{equation}
which is positive and vanishes if and only if the network is fully undirected. As before, the numerical value that is measured has no deep meaning in itself. In order to be interpreted, it must be compared to the value computed for a random network of the same size. Therefore we employ the same procedure as the one previously used to test the community structure of $\left\{ {\cal C}_{1}, {\cal C}_{2}, {\cal C}_{3} \right\}$ (fig.\ref{fig5}) or the rich club structure property (fig.\ref{fig12}), except that it is not the oscillators that are randomly grouped together, but the weights $W$ that are randomly distributed to the links $i \leftarrow j$. The results are shown in fig.\ref{fig13}. Again the convergence to the PDF is very fast and we do check that there is almost no difference between the PDF for $10^5$, $10^6$ and $10^7$ samples. The ASY value of the adapted network is more than $44$ standard deviations below the mean value, which completely rules out the possibility that such a symmetrical network was obtained just by chance. Therefore there is a very clear tendency for the adapted network to become symmetrical.

\subsection{Principal laplacian eigenvectors and natural frequency distribution}
The optimal synchronization conditions of a network of non-identical Kuramoto oscillators has been theoretically derived in \cite{Skardal2014, Skardal2016}. This is an important result because it makes the analitycal link between the network topology and the natural frequency distribution. Although the original result was derived for a directed network with $W_{ij}=K A_{ij}$ and $A_{ij} \in [0,1]$,  it can easily be generalized to the case of eq.(\ref{KuramotoEquation}):
\begin{equation}
 R_{K} \simeq 1-{{\alpha}\over{N}} \displaystyle{\sum_{j=2}^{N}} {{{\big <} u^{j},\tilde{\omega} {\big >}^2}\over{\sigma_{j}^2}}
 \label{RKexpression}
\end{equation}
where $\alpha>0$, $\tilde{\omega} =\left[ \omega_{1},\omega_{2}...\omega_{N}\right]$. $u^{j}$ and $\sigma_{j}$ are related to the singular value decomposition of the Laplacian matrix $L$
\begin{equation}
L_{ij}=\delta_{ij} {\cal M}_{in}(i)-W_{ij}= {\cal U}  \Sigma  {\cal V}^{T}
\label{Laplacian}
\end{equation}
where $R_{K}$ is the usual Kuramoto's order parameter, $ {\cal U}$ and  ${\cal V}$ are unitary matrices, and $\Sigma $ is diagonal. $u^{j}$ and $\sigma_{j}$ occurring in eq.(\ref{RKexpression}) are respectively the jth column of the matrix and the jth element of the diagonal. $\sigma_{j}$ are real and positive. It is usual and practical to rename the eigenvalues so that they appear in ascending order $\sigma_{1}=0 \le \sigma_{2}...\le \sigma_{m}$. $\sigma_{1}$ is vanishing because of the phase invariance $\theta_{j}\longrightarrow \theta_{j}+\phi$ of eq.\ref{KuramotoEquation}.  

In their study \cite{Skardal2014, Skardal2016}, the authors consider a stationary network and look for the natural frequency distribution that optimizes the synchronization.  As the network is fixed, $u^{j}$ and $\sigma_{j}$  are given, and a high degree of  synchronization, i.e. a high value of  $R_{K}$, is therefore associated with a frequency distribution $\tilde{\omega}$ parallel to $u^{m}$ (and therefore orthogonal to $u^{j<m}$). Numerical simulations confirm their results. As for \cite{Brede2008a,Brede2008b}, they do observe that "the heterogeneity of the in-degree distribution matches the heterogeneity of the frequency distribution". But in contrast, they insist that they observe no correlation with the distribution of outgoing connections mass.

\begin{figure}
\resizebox{0.35\textwidth}{!}{
\includegraphics[]{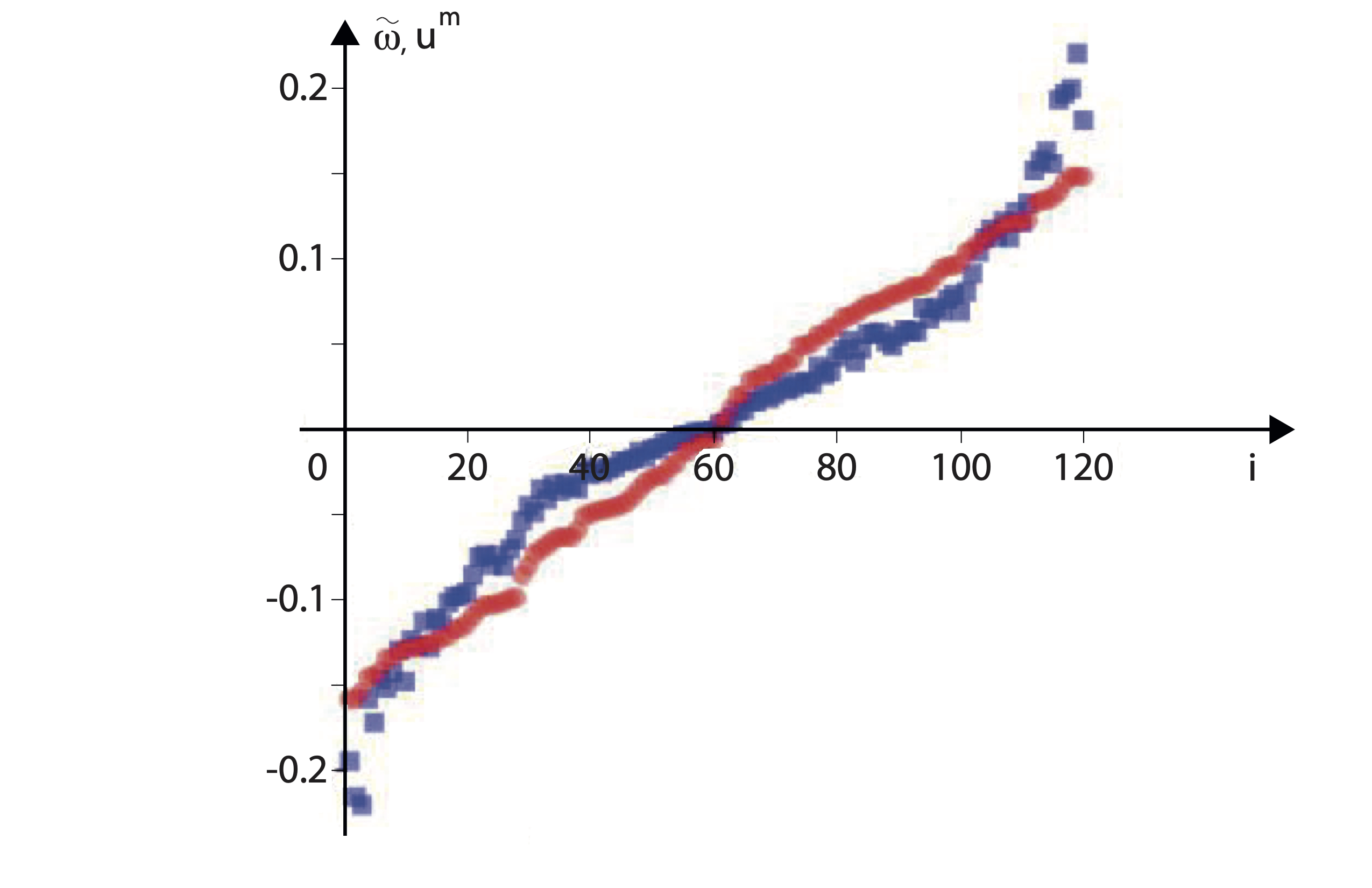}
}
\caption{Comparison between $u^{m}$ (blue squares) and  $\tilde{\omega}$ (red circles) for an optimized Kuramoto's network of $120$ non-identical oscillators. The two vectors have been normalised to $1$. }
\label{fig131}
\end{figure}

\begin{figure}
\resizebox{0.35\textwidth}{!}{
\includegraphics[]{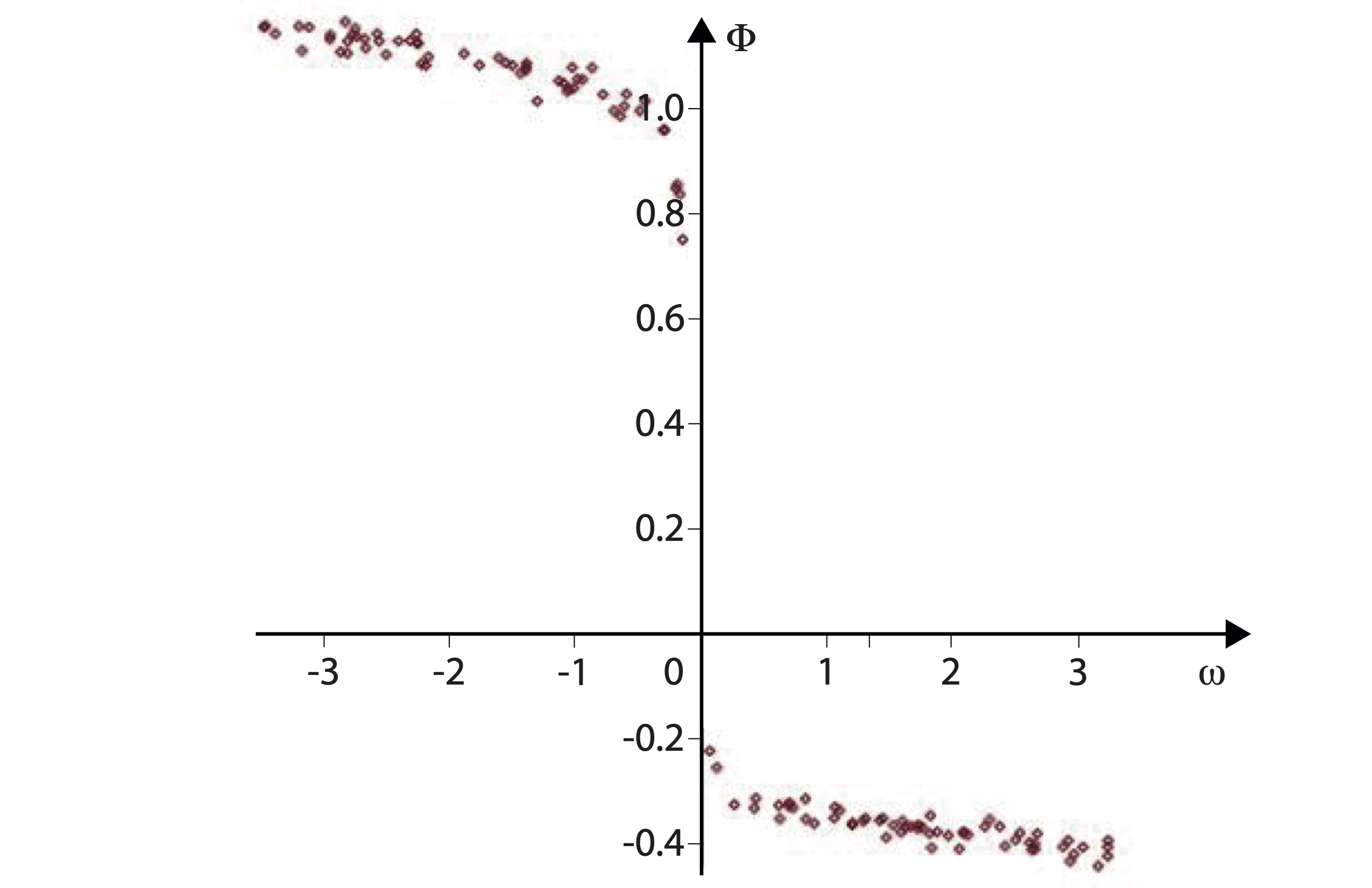}
}
\caption{In the synchronized regime, the phases of the Kuramoto's oscillators satisfy $\theta_{i}=\Omega t +\phi_{i}$, where $\Omega$ is the average pulsation. The plot displays $\phi_{i}$ versus $\omega_{i}$.  Note the phase jump which is incompatible with the quasi-zero phase variation assumption.}
\label{fig132}
\end{figure}

Fig.\ref{fig131} shows the $\tilde{\omega}$ ad  $u^{m}$ vectors corresponding to our own results in the case of the optimization of a network  of $120$ Kuramoto's oscillators. According to the previous theory \cite{Skardal2014, Skardal2016}, the two vectors are indeed almost parallel.  But then, how can we explain our own observation (fig.\ref{fig9}) of a very strong correlation between the distribution of outgoing connections and the distribution of frequencies ? 
The answer is that our problem is different, and in more ways than one.  Firstly, the analytical calculations \cite{Skardal2014, Skardal2016} are carried out under a linearity assumption. It is assumed that, when the oscillators are synchronized, their phases are almost all equal. Technically this means that we can approximate $sin(\theta_{j}-\theta_{i})$ by $(\theta_{j}-\theta_{i})$ in eq.(\ref{KuramotoEquation}). Obviously, this is not what we observe (fig.\ref{fig132}). In our optimization algorithm, we impose a synchronization of the oscillation frequencies, but
as we can see in fig.\ref{fig132}, it does not necessarily lead to phase synchronization. Secondly, the optimization process developed in \cite{Skardal2014, Skardal2016} is of rewiring type, i.e. the total connection mass is conserved, but it is precisely one of the guiding objectives of our algorithm to relax the CTM hypothesis.

\subsection{Network of Ginzburg-Landau oscillators}
\begin{figure}
\resizebox{0.35\textwidth}{!}{
\includegraphics[]{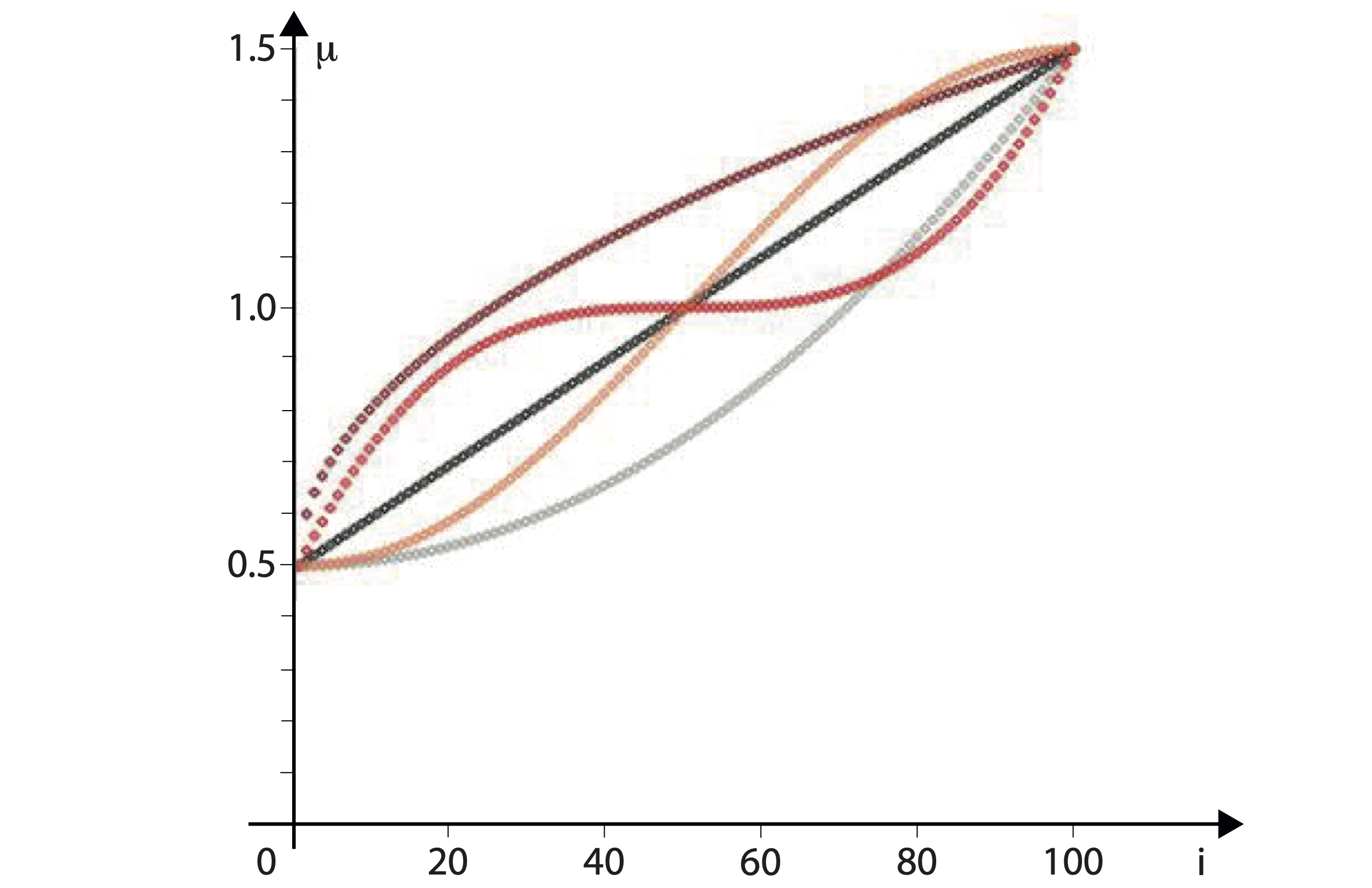}
}
\caption{Various distributions of $\mu_{i}$ used for eq.\ref{GinzburgLandau}. Their analytical expressions are: $\mu_{i}=0.5+f(x_{i})$ with $x_{i}={{i-1}\over{N-1}}$ and $f(x)=x$ (black), $x^2$ (grey), $x^{{1}\over{2}}$ (maroon), $x (4x^2-6x+3)$ (red) and $2.2  x^2+1.2 x^3-4 x^4+1.6 x^5$ (coral). This selection makes it possible to obtain (in order of quotation): a uniform distribution, a surplus of low values, high values, a surplus of intermediate or extreme values.}
\label{fig14}
\end{figure}

For two Kuramoto's phase oscillators with distinct natural frequency $\omega_{1}$ and $\omega_{2}$, the synchronization criterion $W_{12}+W_{21}>\kappa \vert \omega_{1}-\omega_{2} \vert $ is invariant by the exchange  $1 \leftrightarrow 2$. The two oscillators are on an equal footing:  there is neither master nor slave. This is a situation that is not at all generic and most of the time one of two oscillators has a dominant position, for example because its natural oscillation amplitude is the largest. To investigate such a generic effect, we  consider the following network of  simplified Ginzburg-Landau oscillators
\begin{equation}
\partial_{t} A_{i}=\mu_{i} A_{i}-(1+i d) \vert A_{i} \vert ^2 A_{i} + \displaystyle{\sum_{j} W_{ij} A_{j}}
\label{GinzburgLandau}
\end{equation}
where $A_{i}=R_{i} e^{i \theta_{i}}$ stands for the amplitude and phase of oscillator $i$, the $\mu_{i}$ are real, positive and distributed accordingly to fig.\ref{fig14}, $W_{ij}$ are the network real and positive connection weights whose values have to be adapted to reach frequency synchronization (i.e. $\sum_{i}\sum_{j} \vert \partial_{t} \theta_{j}-\partial_{t} \theta_{i} \vert=0$). $d$ stands for the non linear renormalisation of the natural frequency with the oscillation amplitude. Note that the previous form eq.(\ref{GinzburgLandau}) is not the most generic one (a priori all the coefficients could be complex, \cite{Kramer}) but that we limit ourselves to this simpler situation for the sake of clarity. Even within this limit, the problem remains complicated: in absence of coupling, existence of solutions $A_{i}=\sqrt{\mu_{i}} e^{i (\mu_{i} t+\psi)}$ shows that not only the amplitudes of the oscillation are inhomogeneous, but also their natural frequencies.

\begin{figure}
\resizebox{0.35\textwidth}{!}{
\includegraphics[]{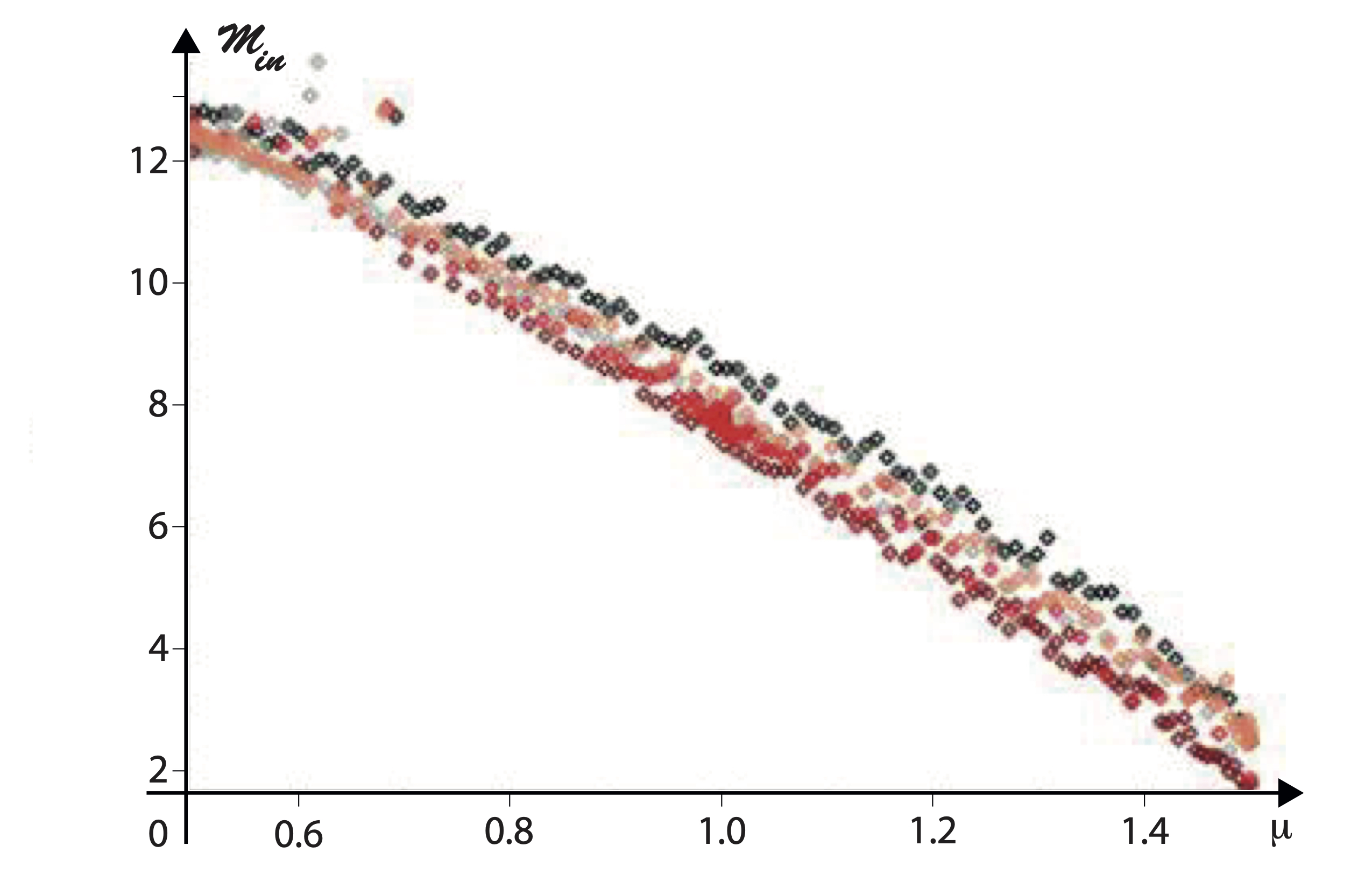}
}
\caption{$N = 100$ Ginzburg-Landau oscillators. Plots of the incoming connection mass ${\cal M}_{in}(i)$ versus $\mu_{i}$ for the distributions described in fig.\ref{fig14}, with the same color code.}
\label{fig15}
\end{figure}

\begin{figure}
\resizebox{0.35\textwidth}{!}{
\includegraphics[]{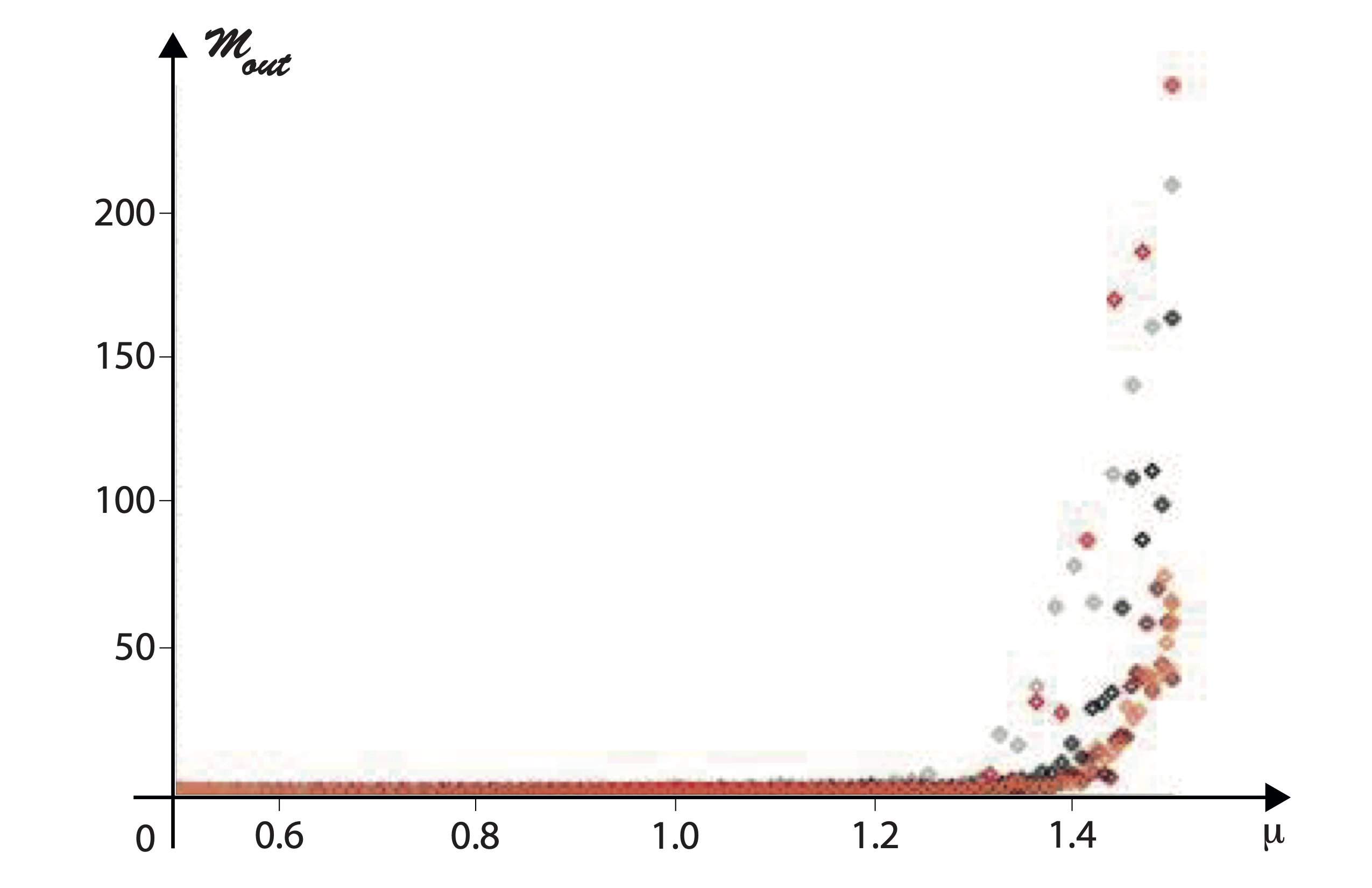}
}
\caption{$N = 100$  Ginzburg-Landau oscillators. Plots of the outgoing connection mass ${\cal M}_{out}(i)$ versus $\mu_{i}$ for the distributions described in fig.\ref{fig14}, with the same color code.}
\label{fig16}
\end{figure}
As for non-identical Kuramoto's oscillators, we do observe, after optimization, a very strong correlation between the distribution of $\mu_{i}$ and that of the masses of incoming (respectively outgoing) connections (fig.\ref{fig15} and respectively fig.\ref{fig16}), attesting a good adaptation of the network to oscillators inhomogeneities. The oscillators with the strongest $\mu_{i}$ are those that emit the most links and receive the least. Introducing
\begin{equation}
\begin{array}{lll}
R_{-}=\{ i \in [1,N],& \mu_{i}&<1.20 \}
\cr
R_{+}=\{ i \in [1,N], 1.20\le &\mu_{i}& \}
\end{array}
\end{equation}
we compute that the relative mass of the connections from $R_{+}$ to $R_{-}$ 
\begin{equation}
{\cal I}={{1}\over{{\cal M}_{tot}}} \displaystyle{\sum_{j \in R_{+}} \sum_{i \in R_{-}} W_{ij}}
\end{equation}
exceeds the mean value by more than $11.7$ (respectively: $25.8$, $ 5.9$, $29.1$, $11.6$)  standard deviations.
Dealing with the asymmetry coefficient (eq.\ref{ASYDefinition}), we find that it exceeds the mean value by more than $3.7$ (respectively $2.9$, $4.7$, $2.0$, $6.3$) standard deviations, which reflects a tendency towards asymmetry. Note the huge difference between Kuramoto and Ginzburg-Landau networks of oscillators:  for the former the asymmetry coefficient was strongly {\it below} the mean value, for the latter, strongly above the mean value. All of these measures show the existence of master-slave relationships on the scale of the entire network.

\begin{figure}
\resizebox{0.35\textwidth}{!}{
\includegraphics[]{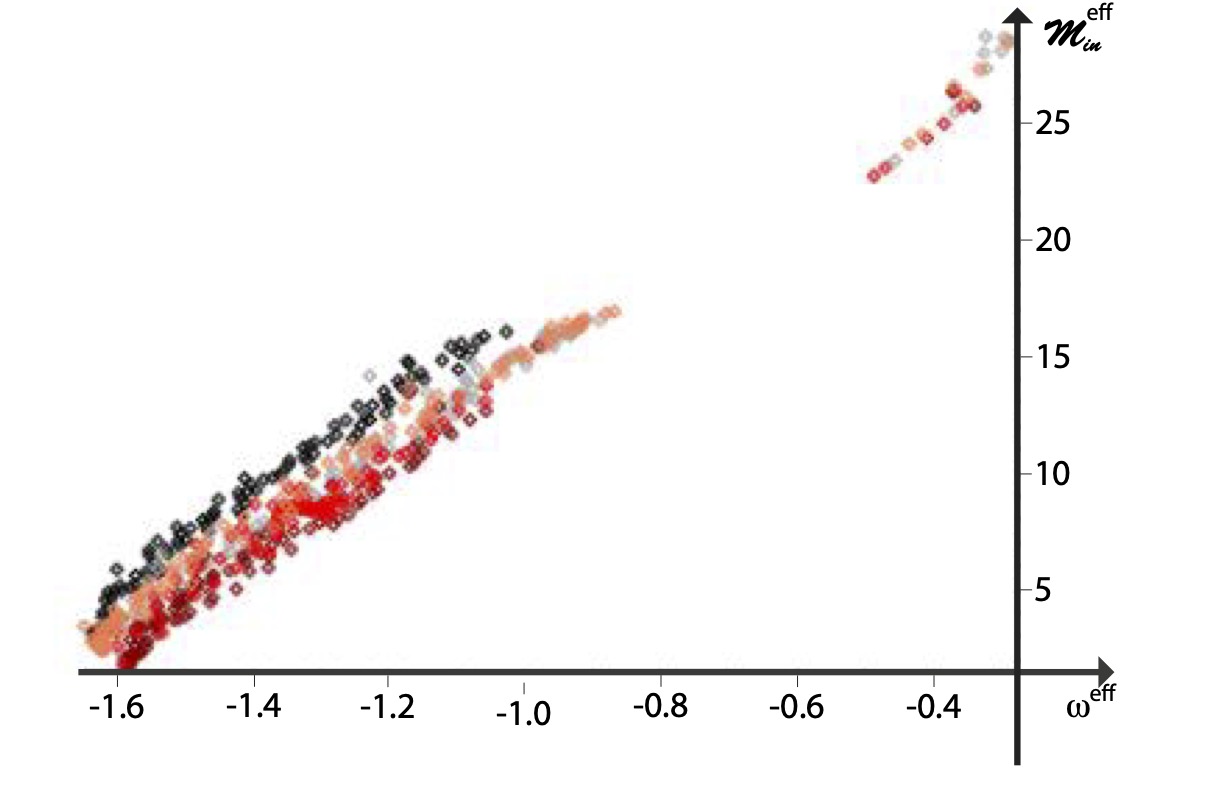}
}
\caption{$N = 100$  Ginzburg-Landau oscillators. Plots of the effective incoming connection mass ${\cal M}_{in}^{\rm{eff}}(i)$ versus $\omega_{i}^{\rm{eff}}$ for the distributions described in fig.\ref{fig14}, with the same color code.}
\label{fig19}
\end{figure}

\begin{figure}
\resizebox{0.35\textwidth}{!}{
\includegraphics[]{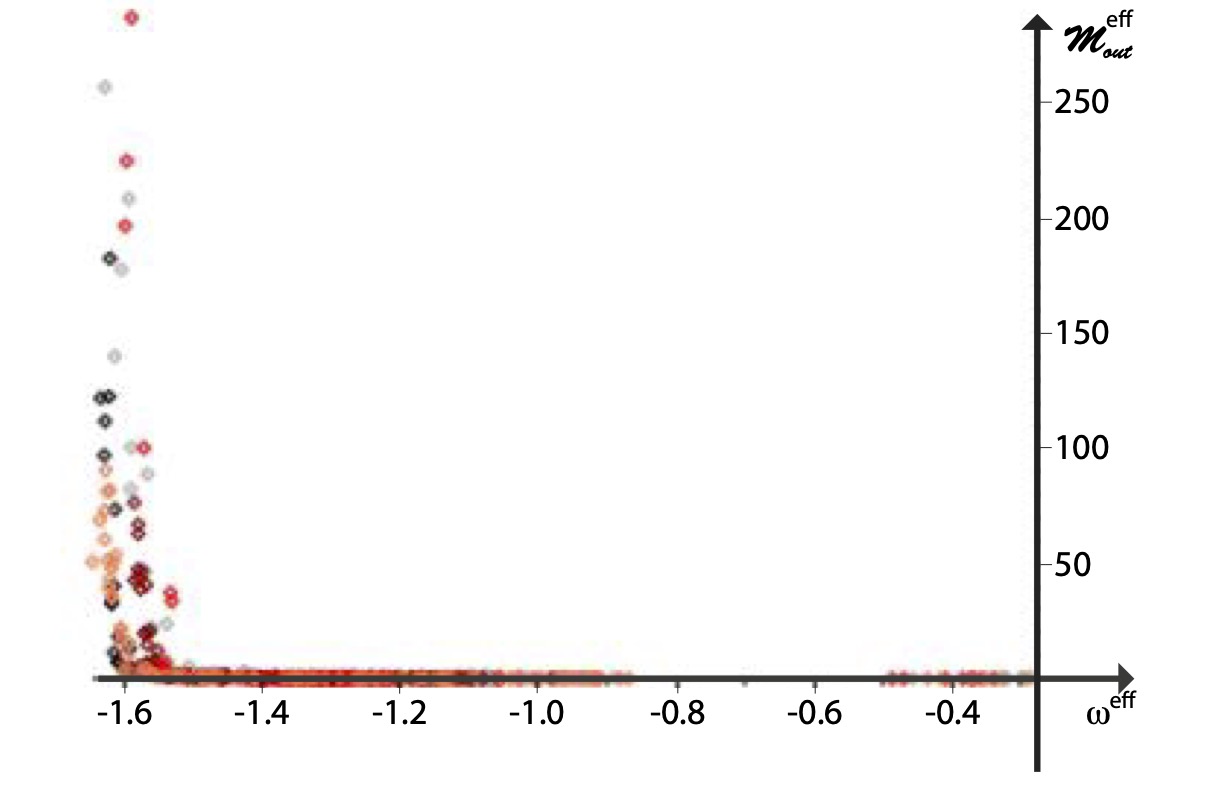}
}
\caption{$N = 100$  Ginzburg-Landau oscillators. Plots of the effective outgoing connection mass ${\cal M}_{out}^{\rm{eff}}(i)$ versus $\omega_{i}^{\rm{eff}}$ for the distributions described in fig.\ref{fig14}, with the same color code.}
\label{fig20}
\end{figure}

It is important to note that in fig.(\ref{fig15}) and fig.(\ref{fig16}), the incoming and outgoing connection masses are expressed versus $ \mu_{i} $ and not versus their natural oscillation frequency  $\omega_{i}$ as in fig.(\ref{fig10}) and fig.(\ref{fig11}). Therefore a direct comparison of the results (i.e. the shape of the curves) is impossible as it stands: the following additional analysis is missing. Substituting $A_{i}=R_{i} e^{i \theta_{i}}$ in eq.(\ref{GinzburgLandau}) leads to
\begin{equation}
\left\{
\begin{array}{ll}
{{\partial R_{i}}\over{\partial t}} &=\mu_{i} R_{i} -R_{i}^3+\displaystyle{\sum_{j}} W_{ij} R_{j}  cos \left( \theta_{j}-\theta_{i} \right)
\cr
{{\partial \theta_{i}}\over{\partial t}} &= -d R_{i}^2+ \displaystyle{\sum_{j}} W_{ij} {{R_{j}}\over{R_{i}}} sin\left(\theta_{j}-\theta_{i}\right)
\end{array}
\right.
\label{decomp}
\end{equation}
where the second equation can be rewritten as an effective Kuramoto's phase equation
\begin{equation}
{{\partial \theta_{i}}\over{\partial t}} =\omega_{i}^{\rm{eff}} + \displaystyle{\sum_{j}} W_{ij} ^{\rm{eff}} sin\left(\theta_{j}-\theta_{i}\right)
\end{equation}
with
\begin{equation}
\begin{array}{lcr}
\omega_{i}^{\rm{eff}}=-d R_{i}^2
& \qquad & W_{ij}^{\rm{eff}}=W_{ij} {{R_{j}}\over{R_{i}}}
\end{array}
\end{equation}
Finally, as the numerical values of the amplitudes $R_{i}$ are directly accessible during the simulation, we can then plot $M_ {in} ^ {\rm{eff}}(i)=\displaystyle{\sum_{j} W_{ij}^{\rm{eff}}}$ and $M_ {out} ^ {\rm{eff}}(i)=\displaystyle{\sum_{j} W_{ji}^{\rm{eff}}}$ versus $\omega_{i}^{\rm{eff}}$ (fig.\ref{fig19} and \ref{fig20}). The corresponding results are in sharp disagreement with the predictions of the analysis of the dynamics reduced to phase equations. This is not really surprising and has already been reported many times in the literature \cite{Kramer}, and this is not even in contradiction with the theorems of dynamic systems (reduction to the central manifold, averaging, adiabatic elimination) which only validate the description in terms of phase over long, but finite times.

\section{Discussion and conclusion}
Scale free networks are characterized by a power law distribution of the incoming connections ${\cal M}_{in}$. It is a measurement that we would have liked to be able to perform, but our computing power does not allow us to explore large networks. At best we have a sample of 120 incoming masses, which is far from being sufficient to calculate a reliable histogram.

Small  world networks are characterised by both a high clustering coefficient and a small diameter.  While we can find in the literature a generalization of the clustering coefficient to weighted and directed networks \cite{Fagiolo2007}, the notion of distance is much more difficult to deal with. Indeed, the distance between two oscillators is expected to be "small"  when the connections between them succeed in imposing their synchronization, and "large" when they do not. Therefore a good candidate for the distance between 2 oscillators has to be related to the stability criterion of the synchronized solution, which depends on the ratio between the mass of their connections and their frequency gap. With this rough definition of the distance between two oscillators, we can look at the distances between all pairs of oscillators in a given network, count the number of times this distance is small and study how this number evolves during the adaptation process. We then observe that this number decreases with time, until it reaches about 30\% of the connections. This means that the synchronization of the adapted network is not achieved by an assembly of small distances, but rather by collective effects for which a pairwise distance does not make much sense. 

We have shown that the gluttonous algorithm can indeed achieve synchronization of non-identical Kuramoto oscillator networks at an economic cost. The algorithm is local, does not require the existence of a central controller and is very easily parallelized. The results concerning the economic aspect of the network, its symmetry, the strong correlation between the distribution of frequencies and the distribution of incoming and outgoing connections, the rich club structure, are both interesting and unexpected. Nevertheless, our brief study of Ginzburg-Landau oscillator networks clearly shows a huge sensitivity of the results to the shape of the dynamics of the nodes themselves. Therefore the generalisation of our study to other types of networks must be particularly circumspect.

\label{Conclusion}

\section{Appendix: review of the previous adaptive network of Kuramoto oscillators }
\label{Appendix}
Following the chronological order, we start with \cite{Seliger2002,Timms2014} where the weight adaptative dynamics obey to
\begin{equation}
\partial_{t} W_{i,j} = \epsilon \left( \alpha \, cos(\theta_{i}-\theta_{j}) -W_{i,j} \right)
\end{equation}
such that the coupling coefficient grows fastest for two oscillators that are in phase and decays fastest for out-of-phase oscillators, mimicking the Hebbian rule of learning \cite{Hebb1949}.
Note that the undirected feature is not imposed, but that $W_{ij}$ and $W_{ji}$  evolve to the same asymptotic value. \cite{Timms2014} revisits  \cite{Seliger2002} but including delays in the synchronization process due to finite propagation velocity. The main result is the existence of multistability: synchronized clusters of different sizes and with different phase relationships among oscillators can be stabilized. The presence of delays accentuates these latter issue.

In \cite{Zanette2006} ,  the weights are of the form
\begin{equation}
W_{ij}=r  A_{ij}    {N \over{N_{i}}}
\end{equation}
where $r$ is the coupling strength, $A_{ij}$$=$$A_{ji}$$=$ $0$ or $1$ the adjacent matrix element between $i$ and $j$, and $N_{i}=\sum_{j=1}^{N} A_{ij}$ is the number of neighbours of $i$. Adaptation is obtained as follows: Regularly after a constant time interval, a node $i$ is chosen at random. Its average frequency $\Omega_{i}$ is measured. Then $\triangle_{ij}=\big \vert \Omega_{j}-\Omega_{i}\big\vert$ is computed for all nodes $j \ne i$, and a node $j_{min}^{node}$ which minimizes  $\triangle_{ij}$ amongst all nodes is identified. The node $j_{max}^{neighbours}$ which maximizes $\triangle_{ij}$ amongst all neighbors of $i$ is also detected. Finally the link between $i$ and $j_{max}^{neighbours}$ is replaced by the link between $i$ and $j_{min}^{node}$ only if the latter is not a neighbour of $i$.
Some remarks are in order:
\begin{enumerate}
\item The network is undirected.
\item The total mass of the connection weight is constant.
\item The weights can only take 2 values: $0$ or $r {{N}\over{N_{i}}}$.
\item The maximum value of $W_{ij}$ is limited (to $r {{N}\over{N_{i}}}$).
\item 2 oscillators may have the same average frequency but completely different instantaneous frequencies.
\end{enumerate}
The main results are i) the convergence of the adaptative network toward a small-world structure, ii) the presence of a rich diversity of synchronized oscillator groups associated with network clusters.

In \cite{Zhao2007}, the weights are ruled by:
\begin{equation}
\partial_{t} W_{i,j} = \epsilon \left( \alpha \vert sin(\beta(\theta_{i}-\theta_{j}))\vert -W_{i,j} \right)
\end{equation}
which implies that the coupling coefficient grows stronger for the pair of oscillators which has larger phase incoherence, contrary to \cite{Seliger2002}. This evolution rule mimics the spike-timing dependent plasticity (STDP)  rule observed in neuroscience \cite{Abbot2000}, where the role of relative spike timing is played by the phase of oscillators. The main result is that such bio inspired rule does enhances the synchronization.

In \cite{Brede2008a,Brede2008b} the weights are on the form
\begin{equation}
W_{ij}=W_{ji}=r A_{ij} N
\end{equation}
Two order parameters are considered: the usual Kuramoto order parameter $R$ in \cite{Brede2008a}  and other one $R_{link}$ in \cite{Brede2008b} which measure the average sum of the absolute value of the local phase difference. The adaptation scheme is the following:
\begin{enumerate}
\item for a given network configuration $W_{ij}$, free evolution for a time interval $T$
\item from $T$ to $2T$, measure of the order parameter
\item a rewired network, where $l$ randomly picked links are swapped to $l$ link vacancies is suggested. Measure of the new order parameter for this rewired network.
\item The new network configuration is accepted only if it gives rise to an improvement of the order parameter
\end{enumerate}
One of the most asset of this approach is the random evolution of the network: there is no attempt to copy nature with Hebbian, anti-Hebbian or STDP evolution rules. Remarkably, despite the absence of bio-inspired rules of evolution, the network does manages to adapt. The other main result is the observation of a positive correlation between the distribution of natural frequencies and the degree of the nodes. However  strong constraints still exist:
\begin{enumerate}
\item The network is undirected.
\item The total mass of the connection weights is constant.
\item The weights can only take 2 values ($0$ or $r$).
\item The maximum value of $W_{ij}$ is limited (to $r$).
\end{enumerate}

In \cite{Assenza2011,AssenzaPRL2011}, the time evolution of the network satisfy to:
\begin{equation}
\partial_{t}W_{ij}=W_{ij} \left( s_{i} \, p_{ij} - \displaystyle{\sum_{k=1}^{N} W_{ik} \, p_{ik} } \right)
\end{equation}
where
\begin{equation}
s_{i}=\displaystyle{\sum_{k=1}^{N} W_{ik}}
\qquad
p_{ij}(t)={\Big \vert} {{1}\over{T}} \displaystyle{\int_{t-T}^{t} e^{i \left( \theta_{j}(x)-\theta_{i}(x)\right) }dx} {\Big \vert}
\end{equation}
where $T$ stands for the process memory length, $s_{i}$ is the mass of the incoming connection, $p_{ij}$ is a measure of the phase synchronization between oscillators $i$ and $j$.
The above equations model the competition between 2 antagonistic mechanisms: Hebbian learning \cite{Hebb1949}  which reinforce those interactions with other correlated units in the graph (also known as homophily in sociology \cite{Cook2001}) and homeostasis which preserves the value of the input strength received by each unit. Indeed, assume that $i$ and $j$ are perfectly synchronized, then $p_{ij}=1$ and
\begin{equation}
\partial_{t}W_{ij} \simeq W_{ij} \left( s_{i} - \displaystyle{\sum_{k=1}^{N} W_{ik} \, p_{ik} } \right)
\end{equation}
but
\begin{equation}
\begin{array}{ll}
p_{ik} \le 1 &\Longrightarrow  \displaystyle{\sum_{k=1}^{N} W_{ik} \, p_{ik} } \le  \underbrace{\displaystyle{\sum_{k=1}^{N} W_{ik}}} _ {s_{i}}
\cr
&\Longrightarrow \partial_{t} W_{ij} \ge 0
\end{array}
\end{equation}
such that the connection weights between synchonized nodes are increased. As for homeostasis, it is automatically included in the fact that $\partial_{t} s_{i}=0$.
The main outcome is the finding that  the competition between these two adaptive principles leads to the emergence of key structural properties observed in real world networks, such as modular and scale free structures, together with a striking enhancement of local synchronization in systems with no global order.

In \cite{Eom2016}, the weights are given by
\begin{equation}
W_{ij}=\lambda N A_{ij}
\end{equation} 
where $\lambda$ is the coupling strength and $A_{ij}=A_{ji}$  the network's adjacency matrix element. Adaptation is achieved through a stochastic time evolution of $A_{ij}(t)$
\begin{equation}
A_{ij}(t)=\left\{
\begin{array}{lllr}
1 & \rm{with} & \rm{probability} &f(i,j,t)
\cr
0 & \rm{with} &  \rm{probability} &1-f(i,j,t)
\end{array}
\right.
\end{equation}
and
\begin{equation}
f(i,j,t)=z {{1+cos(\theta_{j}-\theta_{i})}\over{N}}
\end{equation}
such that two oscillators are more likely to establish a link if they are synchronized. The authors report on an  efficient and concurrent enhancement of
percolation and synchronization through spontaneous self organisation.

In \cite{Papadopoulos2017} the weights are defined as
\begin{equation}
W_{ij}=\alpha N A_{ij}
\end{equation}
with $A_{ij}=A_{ji}$  the adjacent matrix element. At regular times rewiring is invoked as follows: i) first a node $i$ is selected at random. Then 
\begin{equation}
f(i,j)={{1}\over{2}} \left(1+cos(\theta_{j}-\theta_{i})\right)
\end{equation}
is computed for all neighbours $j$ of $i$ and the neighbour $j_{min}^{neighbours}$ which minimizes $f(i,j)$ is identified. Finally the link between $i$ and $j_{min}^{neighbours}$ is broken and a new link is formed with a randomly selected node which is not a neighbour of $i$.
This model displays strong analogy with \cite{Zanette2006,Eom2016}. Note that:
\begin{enumerate}
\item The network is undirected.
\item The total mass of the connection weight is constant.
\item The weights can only take 2 values.
\item The maximum value of $W_{ij}$ is limited.
\end{enumerate}
The observation and description of a spontaneous self organization into structured topologies that support enhanced synchronization dynamics, is the main result of this study.

In \cite{Avalos2018}, the weights satisfy
\begin{equation}
W_{ij}=\sigma_{c} \alpha_{i,j}
\end{equation}
where $\alpha_{i,j}$ depend on time through
\begin{equation}
\left\{
\begin{array}{l}
\partial_{t} \alpha_{ij}=\left( p_{c}-p_{ij} \right) \alpha_{ij} \left( 1-\alpha_{ij} \right)
\cr
p_{ij} = {{1}\over{2}} {\Big \vert} e^{i\theta_{i}}+e^{i\theta_{j}} {\Big \vert}
\end{array}
\right.
\end{equation}
where $p_{ij} \in [0,1] $ measure the synchronization between $i$ and $j$ and $p_{c}$ is a correlation threshold.The dynamics display 2 fix points, $\alpha_{ij}=0$ and $\alpha_{ij}=1$. The farmer is stable when $p_{ij}>p_{c}$ and the latter when  $p_{ij}<p_{c}$, hence mimicking an anti-Hebbian adaptative rule. They found that the emergent networks spontaneously develop the structural conditions to sustain explosive synchronization \cite{Gomez2011}.

\cite{Rentzeperis2020} proposes diffusion based adaptive rewiring as a parsimonious model for activity-dependent reshaping of brain connectivity structure. At each rewiring step, a node $i$ is randomly chosen. Amongst the nodes which are not connected with $i$, the one with the highest heat transfert with $i$, $j_{out}^{max}$, is identified. Then amongst the neighbours of $i$, the one with the lowest heat transfert is selected $j_{in}^{min}$. Finally the connection $(i,j_{in}^{min})$ is substituted with  $(i,j_{out}^{max})$. As a result, where diffusion is intensive, shortcut connections are established, while underused connections are pruned. 
Some remarks:
\begin{enumerate}
\item the network is weighted but not directed
\item The weight distribution is static: it is initially chosen and does not evolves with time. Either normal or log-normal distributions are investigated. The total mass of the connection weight is constant. What is changing with time is the allocation of weights among the connections.
\end{enumerate}


\begin{thebibliography}{120}
\bibitem{PhaseDescription} The theorems of the type Reduction to the central manifold or of the type: Averaging, assert that the global dynamics can be reduced to the only description of critical variables (here the phase). In physicist terms, we speak of adiabatic elimination of amplitude variables. These theorems are proved for long times, but not infinite. In the case of Ginzburg-Landau's equations, numerous studies \cite{Kramer} show that there are regimes of parameters for which the dynamics can very quickly be left out of the scope of phase description.
\end{thebibliography}
\end{document}